\documentclass[twocolumn,twoside,superscriptaddress,longbibliography]{revtex4-2}
\usepackage[latin9]{inputenc}
\usepackage{color}
\usepackage{graphicx}
\usepackage{amsmath}
\usepackage{amssymb}
\usepackage[version=3]{mhchem}
\usepackage{float}
\usepackage{soul}
\usepackage{changes}
\usepackage{mathrsfs}
\usepackage{comment}
\usepackage{multirow}
\usepackage{xcolor}

\definecolor{kc}{rgb}{0.6,0,0.6}

\begin{document}

\title{\emph{Ab-initio} predictions of spin relaxation, dephasing and diffusion in solids}
\setcounter{page}{1}
\date{\today}
\author{Junqing Xu}
\affiliation{Department of Physics, Hefei University of Technology, Hefei, Anhui, China}
\affiliation{Department of Chemistry and Biochemistry, University of California, Santa Cruz, CA 95064, USA}
\author{Yuan Ping}
\email{yping3@wisc.edu}
\affiliation{Department of Materials Science and Engineering, University of Wisconsin-Madison, WI, 53706, USA}
\affiliation{Department of Physics, University of California, Santa Cruz, CA, 95064, USA}
\begin{abstract}
Spin relaxation, dephasing and diffusion are at the heart of spin-based information technology.
Accurate theoretical approaches to simulate spin lifetimes ($\tau_{s}$), determining how fast the spin polarization and phase information will be lost, 
are important to the understandings of underlying mechanism of these spin processes, and 
invaluable to search for promising candidates of spintronic materials.
Recently, we develop a first-principles real-time density-matrix (FPDM)
approach to simulate spin dynamics for general solid-state
systems. Through the complete first-principles' descriptions of light-matter interaction and scattering processes
including electron-phonon, electron-impurity and electron-electron
scatterings with self-consistent spin-orbit coupling, as well as \emph{ab initio}
Land\'e $g$-factor, our method can predict $\tau_{s}$ at various conditions as a function of carrier density and temperature, under electric and magnetic fields. 
By employing this method, we successfully reproduce experimental results
of disparate materials and identify the key factors affecting spin
relaxation, dephasing, and diffusion in different materials. Specifically, we predict
that germanene has long $\tau_{s}$ ($\sim$100 ns at 50 K), a giant
spin lifetime anisotropy and spin-valley locking effect under electric fields, making it advantageous for spin-valleytronic applications.
Based on our theoretical derivations and \emph{ab initio} simulations,
we propose a new useful electronic quantity, named spin-flip angle
$\theta^{\uparrow\downarrow}$, for the understanding of spin relaxation through intervalley spin-flip scattering processes. Our method can be further applied to other emerging materials and extended to simulate exciton spin dynamics and steady-state photocurrents due to photogalvanic effect.
\end{abstract}
\maketitle

\section{Introduction}

In last two decades, spintronics in unconventional semiconductors
and metals, including two-dimensional materials and their heterostructures\citep{avsar2020colloquium,sierra2021van},
topological and magnetic materials\citep{garcia2020canted,vsmejkal2018topological},
hybrid perovskites\citep{kim2021chiral,zhang2022room} etc., have drawn significant interests owing to its unprecedented potentials in microelectronics and next generation low-power electronics. Spin, as a pure quantum mechanical object, is the fundamental information carrier instead of charge, with much less energy dissipation. Ideally one wants such information preserved as long as possible for stable manipulation.  Therefore, understanding how spins relax and transport is of central importance in spintronics. 

Spin is an unconserved quantity in solids due to its coupling with other quantities, such as electron orbital. Therefore, after excess spins
being generated, spin can loose its polarization (relaxation) and phase (dephasing) due to coupling with environment. 
One critical parameter describing the timescale of such processes is spin lifetime
$\tau_{s}$ including T$_1$ (relaxation) and T$_2$ (dephasing), which is often required to be sufficiently long for stable
detection and manipulation of spin. Accurate and reliable theoretical
approaches to simulate $\tau_{s}$ are demanded for the detailed understandings
of spin dynamics and transport phenomena, and designing new spintronics
materials and devices.

Previously, methods based on model Hamiltonian with empirical parameters\citep{vzutic2004spintronics,wu2010spin,avsar2020colloquium}
were extensively employed for simulation of spin relaxation, dephasing, and
diffusion in solids. While these methods provide some mechanistic
insight, they do not serve as predictive tools for the design of new
materials and may sometimes lead to qualitatively incorrect predictions
due to the use of simplified electronic structures and interactions.

On the other hand, the existing first-principles methodology for $\tau_{s}$
has been mostly based on Fermi's Golden rule\citep{restrepo2012full,fedorov2013impact}
considering spin-flip transitions, which is only applicable to systems
with inversion symmetry or high spin polarization, not suitable to
lots of materials with broken inversion symmetry, promising for quantum computing and spintronics
applications~\citep{vzutic2004spintronics,avsar2020colloquium}.
Other first-principles techniques like Time-Dependent Density Functional
Theory (TDDFT)\citep{marques2012fundamentals} are challenging for
crystalline systems due to high computational cost for describing
phonon relaxations that require large supercells. More importantly,
long simulation time over nanoseconds often required by spin relaxation
is a major difficulty for TDDFT, which is only practical for tens
to a few hundred fs. While spin dynamics based on TDDFT
has been recently performed for ultrafast demagnetization of magnetic
systems within tens of fs\citep{chen2019role,acharya2020ultrafast,krieger2015laser},
the intrinsic time scale and supercell limitations mentioned above
remain. A recent development of spin phonon relaxation based on spin-spin correlation function may be a promising pathway, where its applicability for various spin relaxation and dephasing pathways remains to be explored~\cite{Park2022}. 

To overcome these challenges, we recently developed a first-principles
real-time density-matrix (FPDM) approach\citep{xu2021ab,xu2020spin,xu2023spin}
to simulate spin dynamics and pump-probe Kerr-rotation for general solid-state systems.
The approach is free from empirical parameters and is thus of great
predictive power. Through the complete first-principles descriptions
of light-matter interaction and scattering
processes including electron-phonon, electron-impurity and electron-electron
scatterings with self-consistent spin-orbit coupling (SOC) and Land\'e
$g$-factor, our method can predict $\tau_{s}$ as a function of temperature and carrier density, under electric and magnetic fields. 
The method was applied to disparate materials, including semiconductors and metals, with and without inversion symmetry, in good agreement
with experimental results\citep{xu2020spin,xu2021ab,xu2021giant,habib2022electric,xu2023spin,xu2023substrate}.

In this article, we briefly introduce the theory and implementation
of our method, present its applications and show its predictive power
using a two-dimensional Dirac material - Germanene as a showcase.
Through detailed theoretical analysis, we show how \emph{ab-initio}
simulations improve our understandings of spin relaxation mechanisms
and are used to identify the key quantities and/or factors to spin
relaxation, dephasing, and diffusion. At the end, we discuss how our
FPDM approach can be generalized/extended to simulate spin dynamics
of excitons, instead of free carriers, and transport properties such
as photocurrent and spin currents in broken inversion systems (photogalvanic effect). 

\section{Theory}

\subsection{Density-matrix (DM) master equation}

\subsubsection{Quantum master equation}

To provide a general formulation of quantum dynamics in solid-state
materials, we start from the Liouville-von Neumann equation in the
Schr{\"o}dinger picture, 
%
\begin{align}
\frac{d\rho\left(t\right)}{dt} & =-\frac{i}{\text{\ensuremath{\hbar}}}[H,\rho\left(t\right)],\label{eq:Liouville}\\
H & =H_{0}+H',\label{eq:H}
\end{align}

where $H$, $H_{0}$, and $H'$ are total, unperturbed and perturbation Hamiltonians respectively. In this work,
\begin{align}
H_{0}= & H_{e}+H_{\mathrm{ph}}+H_{\mathrm{photon}},\\
H'= & H_{\mathrm{e-light}}+H_{\mathrm{e-ph}}+H_{\mathrm{e-i}}+H_{\mathrm{e-e}},
\end{align}

where $H_{e}$, $H_{\mathrm{ph}}$ and $H_{\mathrm{photon}}$ are
unperturbed single-particle electronic, phonon, and photon Hamiltonian respectively.
$H_{\mathrm{e-light}}$ is the light-matter interaction term. 
$H_{\mathrm{e-ph}}$,
$H_{\mathrm{e-i}}$ and $H_{\mathrm{e-e}}$ describe the electron-phonon
(e-ph), electron-impurity (e-i), and electron-electron (e-e) interaction, respectively.

In practice, the many-body density matrix in Eq.~\ref{eq:Liouville}
is reduced to one-particle density matrix for electrons, where the environmental degree
of freedom is traced out\citep{rossi2002theory}, with a proper truncated BBGKY hierarchy~\citep{Bonitz,Axt1994-vv}. The total rate
of change of the electronic DM $\rho$ is separated into terms related
to different parts of Hamiltonian,
\begin{align}
\frac{d\rho}{dt}= & \frac{d\rho}{dt}|_{\mathrm{coh}}+\frac{d\rho}{dt}|_{\mathrm{scatt}},\label{eq:dynamics}
\end{align}

where $\left[H,\rho\right]=H\rho-\rho H$. $\frac{d\rho}{dt}|_{\mathrm{coh}}$
describes the coherent dynamics of electrons including the free-particle dynamics, the field-induced dynamics, etc. 
$\frac{d\rho}{dt}|_{\mathrm{scatt}}$
captures the scattering between electrons and other particles.
Their detailed forms are given in the subsections below.

To obtain Eq. \ref{eq:dynamics} which involves only the dynamics
of electrons or the electronic subsystem, we have assumed the environmental
subsystem is characterized by a huge number of degrees of freedom
and is not perturbed by the electronic subsystem; in this work it
means there is no dynamics of phonons and photons. Formally this is the prerequisite for the Markovian approximation. 
This assumption
is valid when the system is not far from initial equilibrium, e.g.,
the excited carrier density is not very high. The inclusion of dynamics of the environment, e.g. phonon degrees
of freedom in the dynamics, or formally non-Markovian approximation by taking into account the memory effect of phonon bath, has been discussed in details in Refs. \citenum{iotti2017phonon,rossi2002theory}.

\subsubsection{Coherent terms and external fields\label{subsec:coherent_dynamics_and_external_fields}}

In general, a coherent term corresponding to a single-particle electronic Hamiltonian $H_{\mathrm{sp}}$ reads
\begin{align}
\frac{d\rho}{dt}|_{\mathrm{sp}}= & -\frac{i}{\text{\ensuremath{\hbar}}}\left[H_{\mathrm{sp}},\rho\right].
\end{align}

In absence of external fields, the coherent dynamics contains only one term
\begin{align}
\frac{d\rho}{dt}|_{\mathrm{free}}= & -\frac{i}{\text{\ensuremath{\hbar}}}\left[H_{e},\rho\right].
\end{align}

$H_{e}$ is the unperturbed electronic Hamiltonian, practically computed at the mean-field level, i.e. from density functional theory (DFT). With the electronic eigen-basis,
\begin{align}
H_{e,kmn}= & \epsilon_{kn}\delta_{mn},
\end{align}

where $k$ is k-point index, $m$ and $n$ are band indices, $\epsilon$
is the single-particle energy, and $\delta_{mn}$ is Kronecker delta function.

Therefore, we have 
\begin{align}
\left(\frac{d\rho}{dt}|_{\mathrm{free}}\right)_{kmn}= & -\frac{i}{\text{\ensuremath{\hbar}}}\left(\epsilon_{km}-\epsilon_{kn}\right)\rho_{kmn}.
\end{align}

We note that the exchange-correlation potential in electronic Hamiltonian could be time-dependent, which will introduce an additional term in coherent dynamics\cite{wu2010spin,marini2013competition}. We will not discuss in detail here.

The inclusion of external fields in the DM approach for solid-state
materials is not trivial and was studied by many theorists since 1950s\citep{kohn1957quantum,krieger1987quantum,ciancio2004gauge,kane2012zener,sekine2017quantum,ventura2017gauge,iafrate2017quantum}.
Here, we consider three spatially homogeneous fields as follows. \\

\paragraph{2.1 A laser field.\label{par:laser}}

In this work, we approximate that the light-mater interaction $H_{\mathrm{e-light}}$ consists of two parts - a semiclassical part $H_{\mathrm{laser}}$ describing electrons moving in a laser field (i.e., an electromagnetic field caused by the incident laser) and a quantum part describing the electron-photon interaction in the vacuum without considering the laser.
The semiclassical part corresponds to light absorption and stimulated emission under a laser field, while the quantum part corresponds to spontaneous emission.
We discuss the coherent term due to the semiclassical part first but discuss the quantum part in the next subsection.

The Hamiltonian of a laser with frequency $\omega_{\mathrm{laser}}$
is approximately\citep{xu2021ab}
\begin{align}
H_{\mathrm{laser},kmn}\left(\omega,t\right)= & \frac{e}{m_{e}}{\bf A}_{\mathrm{laser}}\left(t\right)\cdot{\bf p}_{kmn}\mathrm{exp}\left(i\omega_{\mathrm{laser}}t\right)+H.C.,
\end{align}

where \textit{H.C.} is Hermitian conjugate. ${\bf A}_{\mathrm{laser}}\left(t\right)$
is the amplitude. ${\bf A}_{\mathrm{laser}}\left(t\right)$ is real
(complex) for linearly (circularly) polarized light.
${\bf p}_{kmn}$ is the momentum operator matrix element. Note that to include the contribution from the nonlocal part of the pseudopotentials, momentum operator is computed with commutator $[{\bf r},H_e]=\nabla$.

For a Gaussian pump pulse centered at time $t_{\mathrm{center}}$
with width $\tau_{\mathrm{pump}}$,
\begin{align}
{\bf A}_{\mathrm{laser}}\left(t\right) & ={\bf A}_{\mathrm{laser}}\frac{\mathrm{1}}{\sqrt{\sqrt{\pi}\tau_{\mathrm{pump}}}}\mathrm{exp}\left[-\frac{\left(t-t_{\mathrm{center}}\right)^{2}}{2\tau_{\mathrm{pump}}^{2}}\right].
\end{align}
Note that the corresponding pump power is $I_{\mathrm{laser}}=\omega_{\mathrm{laser}}^{2}|{\bf A}_{\mathrm{laser}}|^{2}/\left(8\pi\alpha\right)$,
where $\alpha$ is fine structure constant.

The corresponding dynamics is
\begin{align}
\frac{d\rho}{dt}|_{\mathrm{laser}}= & -\frac{i}{\text{\ensuremath{\hbar}}}\left[H_{\mathrm{laser}},\rho\right].\label{eq:laser-dynamics}
\end{align}
\\

\paragraph{2.2 A static electric field along a non-periodic direction.}
Such electric field along non-periodic direction $E_{\mathrm{np}}$ is directly included in the DFT calculation and is modeled by a ramp or saw-like potential.
Under such treatment, $H_{e}$ is $E_{\mathrm{np}}$-dependent.
As all terms of the master equation (Eq. \ref{eq:dynamics}) are built on eigensystems from $H_{e}$, they are all $E_{\mathrm{np}}$-dependent.
However, if electric field is applied along the periodic direction, one needs to properly treat the periodic boundary condition (e.g. using the modern theory of polarization~\cite{King1993,Resta1994}).  

\paragraph{2.3 A magnetic field.\label{par:theory-Bfield}}

We describe the effects of an external magnetic field ${\bf B}^{\mathrm{ext}}$
using Zeeman Hamiltonian
\begin{align}
H_{Z,k}\left({\bf B}^{\mathrm{ext}}\right)= & \mu_{B}{\bf B}^{\mathrm{ext}}\cdot\left({\bf L}_{k}+g_{0}{\bf S}_{k}\right),\label{eq:Zeeman}
\end{align}
where $\mu_{B}$ is Bohr magneton; $g_{0}$ is the free-electron $g$-factor;
${\bf S}$ and ${\bf L}$ are the spin and orbital angular momentum
respectively. The simulation of ${\bf L}$ is nontrivial for periodic
systems. With the Bl{\"o}ch basis, the orbital angular momentum reads
\begin{align}
{\bf L}_{k,mn}= & i\left\langle \frac{\partial u_{km}}{\partial\mathrm{{\bf k}}}\right|\times\left(\widehat{H}_{e}\left({\bf B}^{\mathrm{ext}}=0\right)-\overline{\epsilon}_{kmn}\right)\left|\frac{\partial u_{kn}}{\partial\mathrm{{\bf k}}}\right\rangle ,\label{eq:Lmatrix}\\
\overline{\epsilon}_{kmn}= & \frac{\epsilon_{km}+\epsilon_{kn}}{2},\label{eq:Lmatrix2}
\end{align}
where $u$ is the periodic part of the single-particle wavefunction; $\widehat{H}_{e}\left({\bf B}^{\mathrm{ext}}=0\right)$
is the zero-field Hamiltonian operator. Eqs.~\ref{eq:Lmatrix} - \ref{eq:Lmatrix2} can
be proven equivalent to ${\bf L=0.5*\left({\bf r\times{\bf p-{\bf p\times{\bf r}}}}\right)}$
with ${\bf r}$ the position operator. The detailed implementation
of Eq.~\ref{eq:Lmatrix} is described in Ref.~\citenum{Multunas2022}.

There are two ways to consider magnetic-field effects.
The first way is including $H_{Z,k}\left({\bf B}^{\mathrm{ext}}\right)$ in $H_{e}$ perturbatively (instead of self-consistently),
then the new eigensystems can be obtained by diagonalizing $H_{e}\left({\bf B}^{\mathrm{ext}}\neq0\right)$.
The second way is including $H_{Z,k}\left({\bf B}^{\mathrm{ext}}\right)$
in $H'$ and the corresponding coherent dynamics is $-\frac{i}{\text{\ensuremath{\hbar}}}\left[H_{Z,k}\left({\bf B}^{\mathrm{ext}}\right),\rho\right]$.
In practice, the two approaches lead to nearly the same dynamical quantities such as lifetimes, since $H_{Z,k}\left({\bf B}^{\mathrm{ext}}\right)$ is rather weak - e.g. Zeeman splitting under 1 Tesla is of order 0.1 meV for many solid-state systems.
In addition, we note that we do not consider the effect of very strong magnetic field such as the appearance of Landau level in this work. 

\subsubsection{Scattering terms}

The scattering part of quantum master equation can be separated into contributions
from several scattering channels, 
\begin{align}
\frac{d\rho}{dt}|_{\mathrm{scatt}}= & \sum_{c}\frac{d\rho}{dt}|_{c},
\end{align}
where $c$ labels a scattering channel corresponding to an interaction
Hamiltonian $H_{c}$. Under Born-Markov approximation and neglecting
renormalization of single particle energies, in general we have\citep{rosati2014derivation}
\begin{align}
\frac{d\rho_{12}}{dt}|_{c}= & \frac{1}{2}\sum_{345}\left[\begin{array}{c}
\left(I-\rho\right)_{13}P_{32,45}^{c}\rho_{45}\\
-\left(I-\rho\right)_{45}P_{45,13}^{c,*}\rho_{32}
\end{array}\right]+H.C.,\label{eq:scattering_BornMarkov}
\end{align}
where $P^{c}$ is the generalized scattering matrix and 
\textit{H.C.}
denotes Hermitian conjugate. The subindex, e.g., ``1'', is the combined
index of k-point and band. The weights of k points must be considered
when summing over k points.

\paragraph{3.1 Electron-phonon.}

The electron-phonon (e-ph) scattering matrix is given by\citep{rosati2014derivation} 
\begin{align}
P_{1234}^{\mathrm{e-ph}}= & \sum_{q\lambda\pm}A_{13}^{q\lambda\pm}A_{24}^{q\lambda\pm,*},\label{eq:Peph}\\
A_{13}^{q\lambda\pm}= & \sqrt{\frac{2\pi}{\hbar}}g_{13}^{q\lambda\pm}\sqrt{\delta_{\sigma}^{G}\left(\omega_{13}\pm\omega_{q\lambda}\right)}\sqrt{n_{q\lambda}^{\pm}},\label{eq:Aeph}\\
\omega_{13}= & \epsilon_{1}-\epsilon_{3},
\end{align}
where $q$ and $\lambda$ are phonon wavevector and mode, $g^{q\lambda\pm}$
is the e-ph matrix element, resulting from the absorption ($-$) or
emission ($+$) of a phonon, $n_{q\lambda}^{\pm}=n_{q\lambda}+0.5\pm0.5$
in terms of phonon Bose factors $n_{q\lambda}$, and $\delta_{\sigma}^{G}$
represents an energy conserving $\delta$-function broadened to a
Gaussian of width $\sigma$. The explicit derivation of Lindblandian dynamics and scattering matrix can be found in Ref.~\citep{rosati2014derivation}. This form guarantees the positive definition of density matrix. 
$g^{q\lambda\pm}$ is computed fully from first-principles with self-consistent
spin-orbit coupling based on density-functional perturbation theory and Wannier interpolation methods\citep{giustino2017electron}.

\paragraph{3.2 Electron-impurity.\label{par:e-i}}

Similar to the e-ph scattering, we write electron-impurity scattering as 
\begin{align}
P_{1234}^{\mathrm{e-i}}= & A_{13}^{i}A_{24}^{i,*},\label{eq:Pei}\\
A_{13}^{i}= & \sqrt{\frac{2\pi}{\hbar}}g_{13}^{i}\sqrt{\delta_{\sigma}^{G}\left(\omega_{13}\right)}\sqrt{n_{i}V_{\mathrm{cell}}},\\
g_{13}^{i}= & \left\langle 1\right|\Delta V^{i}\left|3\right\rangle ,\\
\Delta V^{i}= & V^{i}-V^{0},
\end{align}
where $n_{i}$ and $V_{\mathrm{cell}}$ are impurity density and unit
cell volume, respectively, $V^{i}$ is the potential of the impurity
system and $V^{0}$ is the potential of the pristine system.
Here we assume impurities are randomly distributed and impurity density
is sufficiently low so that the average distance between neighboring
impurities is sufficiently long with nearly no interactions among impurities.

In this work, $g^{i}$ of neutral and ionized impurities are computed
differently as follows.

For neutral impurities, $V^{i}$ is computed with SOC using a large
supercell including an impurity. To speed up the supercell convergence,
we used the potential alignment method developed in Ref. \citenum{sundararaman2017first}.

For ionized impurities, we approximate $\Delta V^{i}$ as the potential
of point charge and is simply the product of the impurity charge $Z$
and the screened Coulomb potential\citep{jacoboni2010theory}. Such
approximate $\Delta V^{i}$ is accurate in the long-range limit,\citep{jacoboni2010theory} i.e.,
its Fourier transform $\Delta V^{i}\left(q\right)$ is accurate when
$q\rightarrow0$, so that it is most suitable for intravalley relaxation process due to electron-impurity scatterings.
The justification of this approximation is that the long-range part is the most dominant in the electrostatic interaction of ionzied impurities\citep{jacoboni2010theory}.

\paragraph{3.3 Electron-electron interaction.}

The e-e scattering matrix is given by\citep{rosati2014derivation} 
\begin{align}
P_{1234}^{\mathrm{e-e}}= & 2\sum_{56,78}\left(I-\rho\right)_{65}\mathscr{A}_{15,37}\mathscr{A}_{26,48}^{*}\rho_{78},\label{eq:Pee}\\
\mathscr{A}_{1234}= & \frac{1}{2}\left(A_{1234}-A_{1243}\right),\\
A_{1234}= & \frac{1}{2}\sqrt{\frac{2\pi}{\hbar}}\left[g_{1234}^{e-e}(\delta_{\sigma,1234}^{G})^{1/2}+g_{2143}^{e-e}(\delta_{\sigma,2143}^{G})^{1/2}\right],\\
g_{1234}^{\mathrm{e-e}}= & \left\langle 1\left(r\right)\right|\left\langle 2\left(r'\right)\right|V\left(r-r'\right)\left|3\left(r\right)\right\rangle \left|4\left(r'\right)\right\rangle ,
\end{align}
where $V\left(r-r'\right)$ is the screened Coulomb potential and
$\delta_{\sigma,1234}^{G}=\delta_{\sigma}^{G}\left(\epsilon_{1}+\epsilon_{2}-\epsilon_{3}-\epsilon_{4}\right)$
is a Gaussian-broadened energy conservation function. The screening
is described by Random-Phase-Approximation (RPA) dielectric function.
We note that unlike the e-ph and e-i channels, $P^{\mathrm{e-e}}$
is a function of $\rho$ and needs to be updated during time evolution
of $\rho$. This is a clear consequence of the two-particle nature
of e-e scattering. $P^{\mathrm{e-e}}$ can be written as the difference
between a direct term ($P^{\mathrm{e-e},d}$) and an exchange term ($P^{\mathrm{e-e},x}$), 
\begin{align}
P^{\mathrm{e-e}}= & P^{\mathrm{e-e},d}-P^{\mathrm{e-e},x},\\
P^{\mathrm{e-e},d}= & \sum_{56,78}\left(I-\rho\right)_{65}A_{15,37}A_{26,48}^{*}\rho_{78},\\
P^{\mathrm{e-e},x}= & \sum_{56,78}\left(I-\rho\right)_{65}A_{15,37}A_{26,84}^{*}\rho_{78}.
\end{align}
According to Ref. \citenum{rossi2002theory}, the direct term is
expected to dominate the dynamical scattering processes among conduction electrons
or valence electrons, allowing us to neglect the exchange term here.

Currently, we use the static Random Phase Approximation (RPA) dielectric
function for the screening without local-field effects. Given the intraband transition is more important, we take the approximate dielectric function as:
\begin{align}
\varepsilon\left({\bf q}\right)= & \varepsilon_{s}\varepsilon^{\mathrm{intra}}\left({\bf q}\right),
\end{align}

where $\epsilon_{s}$ is the static dielectric constant well-defined for insulators, calculated by Density Functional Perturbation Theory (DFPT)\citep{wu2005systematic}. $\varepsilon_s$ effectively takes into account the interband contribution in the dielectric screening. 
$\varepsilon^{\mathrm{intra}}\left({\bf q}\right)$ is the intraband
contribution which involves only states with free carriers and is
critical for doped semiconductors (with free carriers). We computed it at RPA without local field effect,
\begin{align}
\varepsilon^{\mathrm{intra}}\left({\bf q}\right)= & 1-V^{\mathrm{bare}}\left({\bf q}\right)\sum_{{\bf k}mn}\left(\begin{array}{c}
\frac{f_{{\bf k-q},m}-f_{{\bf k}n}}{\epsilon_{{\bf k-q},m}-\epsilon_{{\bf k}n}}\times\\
|\left\langle u_{{\bf k-q},m}|u_{{\bf k}n}\right\rangle |^{2}
\end{array}\right),\label{eq:dielectric-intra}
\end{align}

where the sum runs over only states having free carriers, e.g., for
a n-doped semiconductor, $m$ and $n$ are conduction band indices.
In the formula above, $f$ is time-dependent non-equilibrium occupation
instead of the equilibrium one $f^{\mathrm{eq}}$. Therefore, if the optical field ${\bf A}_{\mathrm{laser}}\left(t\right)$
is activated, $\varepsilon^{\mathrm{intra}}\left({\bf q}\right)$
will be updated in every time step, as $f$ will differ from $f^{\mathrm{eq}}$
and the difference depends on the excitation density. $V^{\mathrm{bare}}\left({\bf q}\right)=e^{2}/\left(V_{\mathrm{cell}}\varepsilon_{0}|q|^{2}\right)$
is the bare Coulomb potential with $\varepsilon_{0}$ vacuum permittivity.

We then have the matrix elements in reciprocal space,
\begin{align}
g_{1234}^{\mathrm{e-e}}= & V^{\mathrm{scr}}\left({\bf q}_{13}\right)\delta_{{\bf k}_{1}+{\bf k}_{2},{\bf k}_{3}+{\bf k}_{4}}\left\langle u_{1}|u_{3}\right\rangle \left\langle u_{2}|u_{4}\right\rangle ,\\
V^{\mathrm{scr}}\left({\bf q}_{13}\right)= & V^{\mathrm{bare}}\left({\bf q}_{13}\right)/\varepsilon\left({\bf q}_{13}\right),
\end{align}

where $V^{\mathrm{scr}}\left({\bf q}\right)$ is the screened Coulomb
potential and ${\bf q}_{13}={\bf k}_{1}-{\bf k}_{3}$. $\left\langle u_{1}|u_{3}\right\rangle $
is the overlap matrix element between two periodic parts of Bloch wave functions.
\\
\paragraph{(iv) Spontaneous emission.}

As discussed above, in this work, the light-matter interaction consists of a semiclassical part (absorption and simulated emission by a laser field) and a quantum part. The semiclassical part has been discussed above (Sec. \ref{subsec:coherent_dynamics_and_external_fields}). Here we discuss the quantum part, which describes the electron-photon interaction in the vacuum.
Similar to the electron-phonon scattering, we write a similar form for electron-photon interaction (the underlying consideration on positive definition of density matrix is similar)~\citep{rosati2014derivation}. 
\begin{align}
P_{1234}^{\mathrm{\mathrm{sp-em}}}= & \sum_{q\lambda\pm}A_{13}^{\mathrm{photon,}q\lambda\pm}A_{24}^{\mathrm{photon,}q\lambda\pm,*},\label{eq:sp-em}\\
A_{13}^{\mathrm{photon,}q\lambda\pm}= & \sqrt{\frac{2\pi}{\hbar}}g_{13}^{\mathrm{photon,}q\lambda\pm}\sqrt{\delta_{\sigma}^{G}\left(\omega_{13}\pm\omega_{q\lambda}^{\mathrm{photon}}\right)}\\
 & \times\sqrt{n_{q\lambda}^{\mathrm{photon,}\pm}},
\end{align}

where `` +" and `` -" correspond to photon emission and absorption respectively, $\lambda$ is the photon mode, $g_{13}^{\mathrm{photon,}q\lambda\pm}$ is the electron-photon matrix element proportional to ${\bf p}$ matrix element. 
The differences from $P^{\mathrm{e-ph}}$ are: 
(i) As $\omega^{\mathrm{photon}}$ is comparable to the band gap which is much greater than $k_{B}T$, 
$n_{q\lambda}^{\mathrm{photon,}-}\approx0$ and $n_{q\lambda}^{\mathrm{photon,}+}\approx1$. 
This represents that photon absorption is not allowed in vacuum, and every photon emission process emits one photon.
(ii) As photon momentum $q$ is tiny (at long wavelength limit), for $A_{13}^{\mathrm{photon}}$, we have ${\bf k}_{1}\approx\text{{\bf k}}_{3}$.
Therefore,
\begin{align}
A_{13}^{\mathrm{photon,}q\lambda+}= & \sqrt{\frac{2\pi}{\hbar}}g_{13}^{\mathrm{photon,}q\lambda+}\sqrt{\delta_{\sigma}^{G}\left(\omega_{13}+\omega_{q\lambda}^{\mathrm{photon}}\right)}\delta_{{\bf k}_{1}{\bf k}_{3}}.
\end{align}

The detailed form of $P_{1234}^{\mathrm{\mathrm{sp-em}}}$ will be
given in our future work.

\subsection{Spin lifetime and spin diffusion length}

\subsubsection{Spin lifetime: relaxation and dephasing\label{subsec:relaxation-and-dephasing}}

In most spin lifetime $\tau_{s}$ experiments, $\tau_{s}$ of 
ensemble spins are measured. The spin observable of an ensemble is
\begin{align}
S^{\mathrm{tot}}_{i}\left(t\right)= & \mathrm{Tr}\left[s_{i}\rho\left(t\right)\right]=\sum_{k}\sum_{mn}s_{i,kmn}\rho_{knm}\left(t\right),\label{eq:Sobservable}
\end{align}

where $s_{i}$ is spin Pauli matrix in Bl\"och basis along direction $i$. 
For spin dynamics, the total excited or excess spin observable is often more relevant and is defined as,
\begin{align}
S^{\mathrm{ex}}_{i}\left(t\right)= & S^{\mathrm{tot}}_{i}\left(t\right) - S^{\mathrm{eq}}_{i}\left(t\right),
\label{eq:Sexess}
\end{align}

where ``eq'' corresponds to the final equilibrium state.

The time evolution must start at an initial state (at $t\,=\,t_{0}$)
with a net spin i.e. $\delta\rho(t_{0})=\rho(t_{0})-\rho^{\mathrm{eq}}\neq0$
such that $S^{\mathrm{ex}}_{i}\left(t\right)\neq0$. We evolve
$\rho\left(t\right)$ through the quantum master equation Eq. \ref{eq:dynamics}
for a long enough simulation time, typically from a few ps to a few
hundred ns, until the evolution of $S^{\mathrm{ex}}_{i}\left(t\right)$ can be reliably
fitted by 
\begin{align}
S^{\mathrm{ex}}_{i}\left(t\right)= & S^{\mathrm{ex}}_{i}\left(t_{0}\right)\mathrm{exp}\left[-\frac{t-t_{0}}{\tau_{s,i}}\right]\\\nonumber
 & \times\mathrm{cos}\left[\omega\left(t-t_{0}\right)+\phi\right]\label{eq:exp_decay}
\end{align}
to extract the relaxation time, $\tau_{s,i}$. Above, $\omega$ is
oscillation frequency due to Larmor precession.

In Ref. \citenum{xu2021ab}, we have shown that one can
generate the initial spin imbalance by applying a test magnetic field
at $t=-\infty$, allowing the system to equilibrate with a net spin
and then turning it off suddenly at $t_{0}$, in order to measure spin relaxation. 

Historically, spin relaxation time (or longitudinal time) $T_{1}$
and ensemble spin dephasing time (or transverse time) $T_{2}^{*}$
were used to characterize the decay of spin ensemble\citep{wu2010spin,lu2007spin}.
Suppose the spins of the system are polarized along direction ${\bf r}_{0}$, possibly
due to applying a constant external field ${\bf B}_{0}$ along ${\bf r}_{0}$, and suppose the total excess/excited spin is along direction ${\bf r}_{1}$, 
if ${\bf r}_{1}||\text{{\bf r}}_{0}$ (or ${\bf r}_{1}\perp\text{{\bf r}}_{0}$),
$\tau_{s}$ is called $T_{1}$ (or $T_{2}^{*}$). For $T_{2}^{*}$,
the excess spin ${\bf S}^{\mathrm{ex}}\left(t\right)$ precesses with a
frequency proportional to $\left|{\bf B}_{0}\right|$.

The ensemble spin dephasing rate $1/T_{2}^{*}$ consists of  reversible
part and irreversible part. The reversible part may be removed by
the spin echo technique. The irreversible part is called spin dephasing
rate $1/T_{2}$, which must be smaller than $1/T_{2}^{*}$. According
to Ref. \citenum{lu2007spin}, $T_{2}$ may be also defined using
Eq. \ref{eq:exp_decay} without the need of spin echo but instead
of $S^{\mathrm{tot}}_{i}\left(t\right)$, we need another quantity - the sum of individual
spin amplitudes
\begin{align}
S_{i}^{\mathrm{indiv}}= & \sum_{k}\sum_{mn}\left|s_{i,kmn}\rho_{knm}\left(t\right)\right|.
\end{align}

\subsubsection{Optical measurements}

Experimentally, $\tau_{s}$ is often studied through ultrafast magneto-optical
pump-probe measurements\citep{meier2012optical,dean2016ultrafast}
where excess spin is generated by a circularly-polarized pump pulse
and its evolution is detected by probe pulses.

The probe pulse interacts with the material similarly to the pump
pulse, and could be described in exactly the same way in principle.
However, this would require repeating the simulation for several values
of the pump-probe delay. Instead, since the probe is typically chosen
to be of sufficiently low intensity, we use second-order time-dependent
perturbation theory to capture its interaction with the system~\citep{xu2021ab},

\begin{align}
\Delta\rho^{\mathrm{probe}}= & \frac{1}{2}\sum_{345}\left\{ \begin{array}{c}
\left[I-\rho\left(t\right)\right]_{13}P_{32,45}^{\mathrm{probe}}\rho\left(t\right)_{45}\\
-\left[I-\rho\left(t\right)\right]_{45}P_{45,13}^{\mathrm{probe}}\rho\left(t\right)_{32}
\end{array}\right\} +H.C.,\label{eq:drho_probe}\\
P_{1234}^{\mathrm{probe}}= & \sum_{\pm}A_{13}^{\mathrm{probe,\pm}}A_{24}^{\mathrm{probe,\pm,*}},\\
A_{13}^{\mathrm{probe,\pm}}= & \sqrt{\frac{2\pi}{\hbar}}\frac{e}{m_{e}}\left({\bf A}^{\mathrm{probe}}\cdot{\bf p}\right)\sqrt{\delta_{\sigma}^{G}\left(\omega_{13}\pm\omega^{\mathrm{probe}}\right)}
\end{align}
The change of dielectric function $\Delta\epsilon$ between the excited
state and ground state absorption detected by the probe is then 
\begin{align}
\mathrm{Im}\Delta\epsilon= & \frac{2\pi}{\left(\omega^{\mathrm{probe}}\right)^{3}|{\bf A}^{\mathrm{probe}}|^{2}}\mathrm{Tr}\left(H_{0}\Delta\rho^{\mathrm{probe}}\right).
\end{align}
Note that $\Delta\rho^{\mathrm{probe}}$ contains $|{\bf A}^{\mathrm{probe}}|^{2}$
so that $\mathrm{Im}\Delta\epsilon$ is independent of ${\bf A}^{\mathrm{probe}}$.
The $\mathrm{Im}\Delta\epsilon$ above is a functional of $\rho$
according to Eq.~\ref{eq:drho_probe} and is an extension of the
usual independent-particle $\mathrm{Im}\epsilon$ depending on just occupation function
$f$\citep{molina2017ab}. After obtaining $\mathrm{Im}\Delta\epsilon$, the
real part $\mathrm{Re}\Delta\epsilon$ can be obtained from the Krames-Kronig
relation.

By summing up $\Delta\epsilon$ with the dielectric function for ground
state absorption, we can obtain the excited-state $\epsilon$ as inputs
for Kerr and Faraday rotation calculations\citep{mainkar1996first}.
These correspond to the rotations of the polarization plane of a linearly-polarized light, reflected by (Kerr rotation) and transmitted (Faraday rotation)  through 
the material, after a pump excitation with a circularly-polarized
light. Time-Resolved Kerr/Faraday Rotation (TRKR/TRFR) has been widely used to study spin dynamics of solids\citep{kikkawa1998resonant,kimel2001room}.
In a TRKR experiment, a circularly-polarized pump pulse is used to
excite valence electrons to conduction bands. The transitions
approximately satisfy the selection rule of $\Delta m_{j}=\pm1$ for
left and right circularly-polarized pulses, respectively, where $m_{j}$
is secondary total angular momentum. TRKR works by measuring the polarization changes
of reflected light, which qualitatively is proportional
to the small population imbalance of electronic states with different
$m_{j}$.

Specifically, the Kerr rotation angle $\theta_{K}$ is computed with
dielectric functions by 
\begin{align}
\theta_{K}= & \mathrm{Im\frac{\sqrt{\epsilon_{+}}-\sqrt{\epsilon_{-}}}{1-\sqrt{\epsilon_{+}}\sqrt{\epsilon_{-}}}},\label{eq:kerr}
\end{align}
where $\pm$ denotes the left and right circular polarization, respectively.

\subsubsection{The work flow of a spin dynamics simulation}

\begin{figure*}
\includegraphics[scale=0.3]{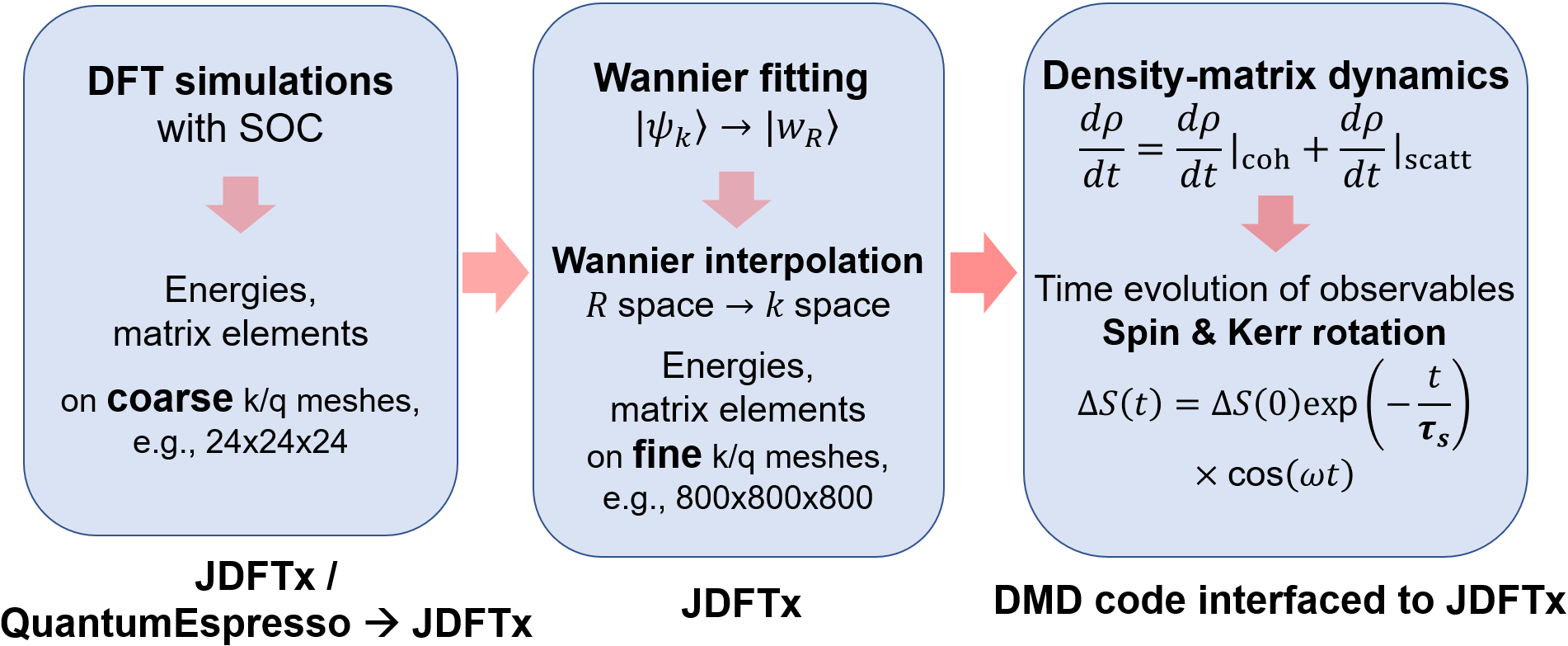}

\caption{The work flow of a spin dynamics simulation. The codes implemented for different
steps are listed and explained in the main text.\label{fig:work-flow}}
\end{figure*}

As shown in Fig. \ref{fig:work-flow}, a spin dynamics simulation
has three main steps:

(i) \textbf{DFT step.} The ground-state electronic structure, phonons,
as well as the e-ph and e-i matrix elements are firstly calculated
using density functional theory (DFT) with relatively coarse $k$
and $q$ meshes in the DFT plane-wave code JDFTx\citep{sundararaman2017jdftx}. The phonon calculation uses the finite difference method with supercells. Alternatively, we can compute the same quantities in QuantumEspresso, with phonon and e-ph couplings calculated by DFPT at the coarse mesh as well. 

(ii) \textbf{Wannier step.} We then transform all quantities from
plane wave basis to maximally localized Wannier function basis\citep{marzari1997maximally},
and interpolate them\citep{PhononAssisted,giustino2017electron,GraphiteHotCarriers,brown2017experimental,NitrideCarriers,TAparameters}
to substantially finer k and q meshes. The Wannier interpolation approach
fully accounts for long-range polar terms (originally from LO-TO splitting) in the e-ph matrix elements and phonon
dispersion relations using the approaches of Ref. \citenum{verdi2015frohlich}
and \citenum{sohier2017breakdown} for the 3D and 2D systems. This part is performed in JDFTx code. 

(iii) \textbf{Dynamics step.} Starting from an initial state with
a spin imbalance, we evolve $\rho\left(t\right)$ through the quantum
master equation of Eq.~\ref{eq:dynamics}. After obtaining spin observable
$S\left(t\right)$ from $\rho\left(t\right)$ (Eq. \ref{eq:Sobservable})
and fitting $S\left(t\right)$ to an exponentially oscillating decay
curve (Eq. \ref{eq:exp_decay}), the decay constant $\tau_{s}$ is
obtained. This part is performed in the DMD code interfacing with the JDFTx code. 

\subsubsection{Carrier lifetime and spin diffusion length}

\paragraph{4.1 Semiclassical limit and carrier lifetime.}

At the semiclassical limit, $\rho$ is replaced by (non-equilibrium)
occupation $f$, then the scattering term originally with a full quantum
description in Eq.~\ref{eq:scattering_BornMarkov} required by DM
dynamics becomes~\citep{xu2021ab}:

\begin{align}
\frac{df_{1}}{dt}|_{c}= & \mathop{\sum_{2\neq1}}\left[\left(1-f_{1}\right)P_{11,22}^{c}f_{2}-\left(1-f_{2}\right)P_{22,11}^{c}f_{1}\right],\label{eq:semiclassical}
\end{align}

using the facts that $P_{11,22}$ is real and ``2=1'' term is zero.
``c'' represent a scattering channel. Note that the weights of k
points must be considered when summing over k points.

Suppose $f$ is perturbed from its equilibrium value by $\delta f$,
i.e., $f=f^{\mathrm{eq}}+\delta f$, then insert $f$ after perturbation
into Eq.~\ref{eq:semiclassical} and linearize it,

\begin{align}
\frac{df_{1}}{dt}|_{c}= & -\mathop{\sum_{2\neq1}}\left[P_{11,22}^{c}f_{2}^{\mathrm{eq}}+\left(1-f_{2}^{\mathrm{eq}}\right)P_{22,11}^{c}\right]\delta f_{1},
\end{align}

using the fact that $\delta P_{11,22}$ is always zero, even for the
e-e scattering.

Define carrier/particle lifetime of state ``1'' $\tau_{p,1}^{c}$
by $\frac{df_{1}}{dt}|_{c}=-\frac{\delta f_{1}}{\tau_{p,1}^{c}}$,
we have

\begin{align}
\frac{1}{\tau_{p,1}^{c}}= & \mathop{\sum_{2\neq1}}\left[P_{11,22}^{c}f_{2}^{\mathrm{eq}}+\left(1-f_{2}^{\mathrm{eq}}\right)P_{22,11}^{c}\right].\label{eq:carrier_lifetime}
\end{align}

The linewidth or the imaginary part of self-energy for the scattering
channel $c$ is related to $\tau_{p}^{c}$ by $\mathrm{Im}\Sigma_{1}^{c}=\hbar/\left(2\tau_{p,1}^{c}\right)$.\\

\paragraph{4.2 Spin diffusion length.}

The DM approach has been widely employed to simulate transport properties\citep{kohn1957quantum,iotti2001nature,watanabe2021chiral}
including spin diffusion length $l_{s}$.\citep{wu2010spin} The implementation
of simulating $l_{s}$ based on FPDM approach for general solid-state
materials is however still under development.

In this work, $l_{s}$ are obtained from first-principles using the
commonly-used relation\citep{vzutic2004spintronics,wu2010spin} based
on the drift-diffusion model,

\begin{align}
l_{s}= & \sqrt{D\tau_{s}},\label{eq:ls}\\
D= & -\mu_{c}n_{c}/\frac{dn_{c}}{d\varepsilon_{\mu_{F}}},
\end{align}
where $D$ is the diffusion coefficient of carriers calculated using the generalized
Einstein relation\citep{kubo1966fluctuation}. 
The underlying assumption here is the diffusion coefficient of carrier being the same as spin, which was shown to be true in graphene-derived systems experimentally\cite{maassen2011comparison}.
$\varepsilon_{\mu_{F}}$
is the electron chemical potential. $n_{c}$ is the carrier density. $\mu_{c}$
is the carrier mobility calculated by solving the linearized Boltzmann
equation in momentum-relaxation-time approximation\citep{ciccarino2018dynamics,gunst2016first,mahan2000many},

\begin{align}
\mu_{c,i}= & \frac{e}{n_{c}V_{u}N_{k}}\sum_{1}\left[f^{\mathrm{eq}}\right]'_{1}v_{1,i}^{2}\tau_{m,1},\label{eq:mobility}
\end{align}
where $i=x,y,z$. $N_{k}$ is the number of k points. $V_{u}$ is
the unit cell volume. $n_{c}$ is electron density. $\left[f^{\mathrm{eq}}\right]'$
is the derivative of the Fermi-Dirac distribution function. $v$ is
the band velocity. $\tau_{m}$ is the momentum relaxation time and
is approximated as\citep{ciccarino2018dynamics,gunst2016first,mahan2000many}
\begin{align}
\tau_{m,1}^{-1}= & \mathop{\sum_{2\neq1}}\left\{ \left[P_{11,22}f_{2}^{\mathrm{eq}}+\left(1-f_{2}^{\mathrm{eq}}\right)P_{22,11}\right]\mathrm{cos}\theta_{12}\right\} ,\label{eq:cos_theta}\\
\mathrm{cos}\theta_{12}= & \frac{{\bf v}_{1}\cdot{\bf v}_{2}}{v_{1}v_{2}},
\end{align}

where ${\bf v}$ is the velocity vector. We find that $\tau_{m}$
is similar to the carrier lifetime $\tau_{p}$ except the angle factor
$\mathrm{cos}\theta_{12}$.

\section{Spin relaxation mechanism and analysis methods for spin relaxation}

Through studying the relations between $\tau_{s}$ and important electronic
quantities, switching on/off certain dynamic processes, or tuning the
key factors affecting spin dynamics, our FPDM method is a powerful technique to determine spin relaxation mechanism and quantitatively predict spin lifetime
in different materials at various conditions.

There are several mechanisms causing spin relaxation and dephasing
of electron carriers in non-magnetic systems\citep{vzutic2004spintronics,wu2010spin}. Among them,
the most important mechanisms are Elliot-Yafet (EY) and Dyakonov-Perel
(DP) mechanisms. EY represents spin relaxation due to spin-flip scattering.
DP is activated when inversion symmetry is broken, which results in
random spin precession between adjacent scattering events. Similar
to DP mechanism, free induction decay (FID) mechanism is caused by
random spin precession but happens when scattering is weak enough.
Below we show how to determine spin relaxation mechanism and analyse
spin relaxation through \emph{ab-initio} simulations.
We note that our FPDM approach provides unified treatment for different mechanisms. We also acknowledge that some approximate models were proposed in the past specific for each mechanism, which was helpful for providing  mechanistic insights and qualitative understanding. We will introduce them in the following sections. 
By comparing FPDM and these approximate models, we can further grasp the detailed physical picture of spin relaxation. 

\subsection{Spin expectation value, spin texture and internal magnetic field}

We first define some important spin-related properties in spin dynamics.

The spin expectation value along $i$ direction is defined as the
diagonal element of spin matrix $s_{i}$, i.e.,
\begin{align}
S_{i,1}^{\mathrm{exp}}= & s_{i,11}.
\end{align}
Note that for degenerate bands, the matrix elements of $s_{i}$ are arbitrary,
thus, we need to diagonalize $s_{i}$ matrix in degenerate subspaces.

The spin texture presents when the spin up and down degeneracy (Kramers pair) is lifted (e.g., due
to broken inversion symmetry), and is the distribution of the spin
expectation value vector ${\bf S}^{\mathrm{exp}}\equiv\left(S_{x}^{\mathrm{exp}},S_{y}^{\mathrm{exp}},S_{z}^{\mathrm{exp}}\right)$.

Suppose originally a system has time-reversal and inversion symmetries,
so that every two bands form a Kramers degenerate pair. Suppose the
${\bf k}$-dependent spin matrix vectors in Bloch basis of the Kramers
degenerate pairs are ${\bf s}_{k}^{0}$ with ${\bf s}\equiv\left(s_{x},s_{y},s_{z}\right)$.
The inversion symmetry broken induces ${\bf k}$-dependent Hamiltonian
terms\citep{vzutic2004spintronics}

\begin{align}
H_{k}^{\mathrm{ISB}}= & \mu_{B}g_{0}{\bf B}_{k}^{\mathrm{in}}\cdot{\bf s}_{k}^{0},\label{eq:HISB}
\end{align}

where ${\bf B}_{k}^{\mathrm{in}}$ is the so-called internal
magnetic fields at presence of SOC. ${\bf B}^{\mathrm{in}}$ splits the degenerate pair
and polarizes the spin along its direction. The definition of ${\bf B}_{k}^{\mathrm{in}}$
is

\begin{align}
{\bf B}_{k}^{\mathrm{in}}\equiv & 2\Delta_{k}{\bf S}_{k}^{\mathrm{exp}}/\left(\mu_{B}g_{0}\right),\label{eq:Bin}
\end{align}

where $\Delta$ is the band splitting energy. From Eq. \ref{eq:Bin},
${\bf S}_{k}^{\mathrm{exp}}||{\bf B}_{k}^{\mathrm{in}}$ (internal magnetic field ${\bf B}_{k}^{\mathrm{in}}$ is along the spin texture direction ${\bf S}_{k}^{\mathrm{exp}}$).

\subsection{Elliot-Yafet (EY) mechanism\label{subsec:EY_mechanism}}

EY mechanism dominates spin relaxation when spin-up and spin-down
are well defined (in absence of spin precession), so that it dominates in two types of systems: (i)
Materials with both time-reversal and spatial inversion symmetries,
e.g., silicon, free-standing graphene in absence of external fields. With such symmetries, every
two bands form a Kramers degenerate pair\citep{vzutic2004spintronics}. 
Therefore, spin-up/down is well defined along an axis ${\bf \widehat{r}}$
by diagonalizing the corresponding spin matrix $s_{{\bf r}}={\bf s}\cdot{\bf \widehat{r}}$
in degenerate subspaces. (ii) Systems with high spin polarization e.g. due to intrinsic magnetization or spin-valley lockng,
e.g., ferromagnets, transition metal dichalcogenides (TMDs). 

\subsubsection{Fermi's golden rule (FGR)}

\paragraph{1.1 FGR with spin-flip transitions}

According to Ref.~\citep{xu2020spin}, if a solid-state system is
close to equilibrium (but not at equilibrium) and its spin relaxation
is dominated by EY mechanism, its free carriers' $\tau_{s}$ due to spin-orbit coupling and e-ph scattering
approximately satisfies (for simplicity the band indices are dropped)
\begin{align}
\tau_{s}^{-1}\propto & \frac{N_{k}^{-2}}{\chi}\sum_{kq\lambda}\left\{ \begin{array}{c}
|g_{k,k-q}^{\uparrow\downarrow,q\lambda}|^{2}n_{q\lambda}f_{k-q}^{\mathrm{eq}}\left(1-f_{k}^{\mathrm{eq}}\right)\\
\delta\left(\epsilon_{k}-\epsilon_{k-q}-\omega_{q\lambda}\right)
\end{array}\right\} ,\label{eq:FGR-1}\\
\chi= & N_{k}^{-1}\sum_{k}f_{k}^{\mathrm{eq}}\left(1-f_{k}^{\mathrm{eq}}\right),\label{eq:FGR-2}
\end{align}

where $g^{\uparrow\downarrow}$ is the spin-flip e-ph matrix element
between two electronic states of opposite spins, $n_{q,\lambda}$ is phonon occupation at momentum $q$ and mode $\lambda$, and $f^{\mathrm{eq}}$ is Fermi-Dirac distribution function. We will further discuss
$g^{\uparrow\downarrow}$ in the next subsection.


According to Eq. \ref{eq:FGR-1} and \ref{eq:FGR-2}, $\tau_{s}^{-1}$
is proportional to $|g_{q}^{\uparrow\downarrow}|^{2}$ and also the
density of the spin-flip transitions. Therefore we propose a temperature
($T$) and chemical potential ($\varepsilon_{\mu_{F}}$) dependent
effective modulus square of the spin-flip e-ph matrix element $\overline{|\widetilde{g}^{\uparrow\downarrow}|^{2}}$
and a scattering density of states $D^{\mathrm{S}}$ as 
\begin{align}
\overline{|\widetilde{g}^{\uparrow\downarrow}|^{2}}= & \frac{\sum_{kq}\mathrm{w}_{k,k-q}\sum_{\lambda}|g_{k,k-q}^{\uparrow\downarrow,q\lambda}|^{2}n_{q\lambda}}{\sum_{kq}\mathrm{w}_{k,k-q}},\label{eq:gsf2}\\
D^{\mathrm{S}}= & \frac{N_{k}^{-2}\sum_{kq}\mathrm{w}_{k,k-q}}{N_{k}^{-1}\sum_{k}f_{k}^{\mathrm{eq}}\left(1-f_{k}^{\mathrm{eq}}\right)},\label{eq:DS}\\
\mathrm{w}_{k,k-q}= & f_{k-q}^{\mathrm{eq}}\left(1-f_{k}^{\mathrm{eq}}\right)\delta\left(\epsilon_{k}-\epsilon_{k-q}-\omega_{c}\right),
\end{align}

where $\omega_{c}$ is the characteristic phonon energy chosen as
the averaged energy of the phonons contributing to spin relaxation,
and $\mathrm{w}_{k,k-q}$ is the weight function. The matrix element
modulus square is weighted by $n_{q\lambda}$ according to Eq. \ref{eq:FGR-1}
and \ref{eq:FGR-2}. This rules out high-frequency phonons at low
$T$ which are not excited. $\mathrm{w}_{k,k-q}$ selects transitions
between states separated by $\omega_{c}$ and around the band edge
or $\varepsilon_{\mu_{F}}$, which are ``more relevant'' transitions
to spin relaxation.

$D^{\mathrm{S}}$ can be regarded as an effective density of spin-flip
e-ph transitions satisfying energy conservation between one state
and its pairs. When $\omega_{c}=0$, we have $D^{\mathrm{S}}=\int d\epsilon\left(-\frac{df^{\mathrm{eq}}}{d\epsilon}\right)D^{2}\left(\epsilon\right)/\int d\epsilon\left(-\frac{df^{\mathrm{eq}}}{d\epsilon}\right)D\left(\epsilon\right)$
with $D\left(\epsilon\right)$ density of electronic states (DOS).
So $D^{\mathrm{S}}$ can be roughly regarded as a weighted-averaged
DOS with weight $\left(-\frac{df^{\mathrm{eq}}}{d\epsilon}\right)D\left(\epsilon\right)$.

With $\overline{|\widetilde{g}^{\uparrow\downarrow}|^{2}}$ and $D^{\mathrm{S}}$,
we have the approximate relation for spin relaxation rate, 
\begin{align}
\tau_{s}^{-1}\propto & \overline{|\widetilde{g}^{\uparrow\downarrow}|^{2}}D^{\mathrm{S}}.\label{eq:taus_approx}
\end{align}

We can similarly define an effective spin conserving matrix element
$\overline{|\widetilde{g}^{\uparrow\uparrow}|^{2}}$ by replacing
$g_{k,k-q}^{\uparrow\downarrow,q\lambda}$ to $g_{k,k-q}^{\uparrow\uparrow,q\lambda}$
in Eq. \ref{eq:gsf2}. Then we have the approximate relation for carrier
relaxation rate due to e-ph scattering,
\begin{align}
\left\langle \tau_{p}^{-1}\right\rangle  & \propto\overline{|\widetilde{g}^{\uparrow\uparrow}|^{2}}D^{\mathrm{S}}.\label{eq:taup_gsc_relation}
\end{align}

Although such formula are not as general and accurate as our FPDM formulation, Eq.~\ref{eq:taus_approx} and ~\ref{eq:taup_gsc_relation} can be used to understand the spin flip and spin conserving e-ph matrix element's relation to spin or carrier relaxation, and one can develop intuition on what type of phonons contribute more to $\overline{|\widetilde{g}^{\uparrow\uparrow}|^{2}}$  or $\overline{|\widetilde{g}^{\uparrow\downarrow}|^{2}}$. As our recent studies show Fr\"{o}lich LO phonon strongly contributes to carrier relaxation in CsPbBr$_3$, but much less important in its spin relaxation due to the spin-conserving nature of long-ranged Fr\"{o}lich electron-phonon coupling~\citep{xu2023spin}. 
\\

\paragraph{1.2 Generalized FGR}

In centrosymmetric systems with strong spin-mixing and band degeneracy, ${\bf S}_{k}^{\mathrm{exp}}$
may have multiple values. For example, valence bands of silicon
at $\Gamma$ are four-fold degenerate and their ${\bf S}_{k}^{\mathrm{exp}}$
are (more precisely, very close to) $\pm1/2$ and $\pm1/6$. In such
cases, spin relaxation may be still driven by EY mechanism but
needs a generalized FGR formula\citep{xu2020spin} beyond spin-flip transition,

\begin{align}
\tau_{s,i}^{-1}\propto & \frac{N_{k}^{-2}}{\chi_{i}}\sum_{12\lambda}\left\{ \begin{array}{c}
\left|\Delta S_{i,12}^{\mathrm{exp}}g_{12}^{q,\lambda}\right|^{2}n_{q\lambda}\\
f_{2}^{\mathrm{eq}}\left(1-f_{1}^{\mathrm{eq}}\right)\delta\left(\epsilon_{1}-\epsilon_{2}-\omega_{q\lambda}\right)
\end{array}\right\} ,\label{eq:GenFGR-1}\\
\chi_{i}= & N_{k}^{-1}\sum_{1}f_{1}^{\mathrm{eq}}\left(1-f_{1}^{\mathrm{eq}}\right)S_{i,1}^{\mathrm{exp},2},\label{eq:GenFGR-2}\\
\Delta S_{i,12}^{\mathrm{exp}}= & S_{i,1}^{\mathrm{exp}}-S_{i,2}^{\mathrm{exp}},\label{eq:GenFGR-3}
\end{align}

where $\Delta S^{\mathrm{exp}}$ is the change of spin expectation
value for a pair of electronic states. Therefore, the transitions
with non-negligible $\Delta S^{\mathrm{exp}}$ can contribute to spin
relaxation. Again, unlike our FPDM method, this is approximate formula, which can be used to understand how e-ph scattering between a pair of states changes spin expectation value as we showcase bulk Si hole spin relaxation in Ref.~\citenum{xu2020spin}.

\subsubsection{Spin-mixing parameter $b^{2}$}

\paragraph{2.1 State-resolved $b^{2}$.}

Suppose the spin of a state ``1'' is highly polarized along $i$
direction. Then in general, the wavefunction of state ``1'' can
be written as $\Psi_{1}\left({\bf r}\right)=a_{i,1}\left({\bf r}\right)\alpha+b_{i,1}\left({\bf r}\right)\beta$,
where $a$ and $b$ are the coefficients of the large and small components
of the wavefunction, and $\alpha$ and $\beta$ are spinors (one up
and one down for direction $i$). Define $a_{i,1}^{2}=\int|a_{i,1}\left({\bf r}\right)|^{2}d{\bf r}$
and $b_{i,1}^{2}=\int|b_{i,1}\left({\bf r}\right)|^{2}d{\bf r}$,
then $a_{i,1}^{2}>b_{i,1}^{2}$ and $b_{i,1}^{2}$ is just spin-mixing
parameter of state ``1'' along direction $i$~\citep{xu2021giant}. 

\begin{align}
a_{i,1}^{2}+b_{i,1}^{2}= & 1,\\
0.5\left(a_{i,1}^{2}-b_{i,1}^{2}\right)= & S_{i,1}^{\mathrm{exp}},
\end{align}
Therefore, 
\begin{align}
b_{i,1}^{2}= & 0.5-S_{i,1}^{\mathrm{exp}}.
\end{align}

\paragraph{2.2 EY relation.}

According to Eq. \ref{eq:FGR-1} and \ref{eq:taup_gsc_relation},
we have
\begin{align}
\tau_{s}^{-1}/\tau_{p}^{-1}\propto & \overline{|\widetilde{g}^{\uparrow\downarrow}|^{2}}/\overline{|\widetilde{g}^{\uparrow\uparrow}|^{2}},
\end{align}
where $\tau_{p}=1/\left\langle \tau_{p}^{-1}\right\rangle $.

As thermal averaging  frequently appears in spin relaxation analysis,
we define $\left\langle A\right\rangle $ as the thermal average of
electronic quantity $A$,

\begin{align}
\left\langle A\right\rangle = & \frac{\sum_{kn}\left(-\left[f^{\mathrm{eq}}\right]'_{kn}\right)A_{kn}}{\sum_{kn}\left(-\left[f^{\mathrm{eq}}\right]'_{kn}\right)},\label{eq:thermal_average}
\end{align}
where $\left[f^{\mathrm{eq}}\right]'$
is the derivative of the Fermi-Dirac distribution function.

According to Refs. \citenum{fabian1998spin,leyland2007oscillatory,restrepo2012full},
$\overline{|\widetilde{g}^{\uparrow\downarrow}|^{2}}/\overline{|\widetilde{g}^{\uparrow\uparrow}|^{2}}\sim\left\langle b^{2}\right\rangle $,
so that
\begin{align}
\tau_{s}^{-1}/\tau_{p}^{-1}\sim & \left\langle b^{2}\right\rangle .\label{eq:taus_b2_relation}
\end{align}
The above is a rough relation and cannot be used to predict $\tau_{s}$
(the error may be several or even ten times). Practically, we use
the following approximate relation
\begin{align}
\tau_{s}^{-1}/\tau_{p}^{-1}= & 4\left\langle b^{2}\right\rangle .
\end{align}
This is called EY relation in this work.

\subsubsection{Spin-flip angle $\theta^{\uparrow\downarrow}$\label{subsec:Spin-flip-angle}}

In our previous paper\citep{xu2023substrate}, we proposed an important
electronic quantity for intervalley spin-flip scattering - the spin-flip
angle $\theta^{\uparrow\downarrow}$ between two electronic states.
For two states $\left(k,n\right)$ and $\left(k',n'\right)$ with
opposite spin directions, $\theta^{\uparrow\downarrow}$ is the angle
between $-{\bf S}_{kn}^{\mathrm{exp}}$ and ${\bf S}_{k'n'}^{\mathrm{exp}}$.
The motivation of examining $\theta^{\uparrow\downarrow}$ is that:
According to Ref. \citep{yafet1963g}, due to time-reversal symmetry,
the spin-flip matrix element of the same band between ${\bf k}$ and
$-{\bf k}$ is exactly zero, so that $g^{\uparrow\downarrow}$ is
zero at lowest order for intervalley transitions between two opposite
valleys (e.g., ${\bf K}$ and ${\bf -K}$). In the first-order perturbation
level, $\left|g^{\uparrow\downarrow}\right|$ between two states is
determined by $\theta^{\uparrow\downarrow}$ between these two states
and proportional to $\left|\sin\left(\theta^{\uparrow\downarrow}/2\right)\right|$.

Suppose (i) the inversion symmetry broken induces ${\bf B}_{k}^{\mathrm{in}}$
(Eq. \ref{eq:Bin}) for a Kramers degenerate pair; (ii) there are two
valleys centered at wavevectors ${\bf Q}$ and $-{\bf Q}$ and (iii)
there are two wavevectors ${\bf k}_{1}$ and ${\bf k}_{2}$ near ${\bf Q}$
and $-{\bf Q}$ respectively. Due to time-reversal symmetry, the directions
of ${\bf B}_{k_{1}}^{\mathrm{in}}$ and ${\bf B}_{k_{2}}^{\mathrm{in}}$
are almost opposite.

We can prove that for a general operator $\widehat{A}$,

\begin{align}
\left|A_{k_{1}k_{2}}^{\uparrow\downarrow}\right|^{2}\approx & \mathrm{sin}^{2}\left(\theta_{k_{1}k_{2}}^{\uparrow\downarrow}/2\right)\left|A_{k_{1}k_{2}}^{\downarrow\downarrow}\right|^{2},\label{eq:spin-filp-relation}
\end{align}

where $A_{k_{1}k_{2}}^{\uparrow\downarrow}$ and $A_{k_{1}k_{2}}^{\downarrow\downarrow}$
are the spin-flip and spin-conserving matrix elements between ${\bf k}_{1}$
and ${\bf k}_{2}$ respectively. We present the detailed derivation in SI Sec. SIII of Ref. \citenum{xu2023substrate}.

From Eq. \ref{eq:spin-filp-relation}, for the intervalley e-ph matrix
elements, we have

\begin{align}
\left|g_{k_{1}k_{2}}^{\uparrow\downarrow}\right|^{2}\approx & \mathrm{sin}^{2}\left(\theta_{k_{1}k_{2}}^{\uparrow\downarrow}/2\right)\left|g_{k_{1}k_{2}}^{\downarrow\downarrow}\right|^{2}.
\end{align}

Finally, similar to Eq. \ref{eq:gsf2}, we propose an effective modulus
square of $\mathrm{sin}^{2}\left(\theta_{k_{1}k_{2}}^{\uparrow\downarrow}/2\right)$,

\begin{align}
\overline{\mathrm{sin}^{2}\left(\theta^{\uparrow\downarrow}/2\right)}= & \frac{\sum_{kq}\mathrm{w}_{k,k-q}\mathrm{sin}^{2}\left(\theta_{k,k-q}^{\uparrow\downarrow}/2\right)}{\sum_{kq}\mathrm{w}_{k,k-q}}.\label{eq:sin2}
\end{align}

We found spin relaxation time (obtained from our FPDM method) linearly proportional to this quantity ($\overline{\mathrm{sin}^{2}\left(\theta^{\uparrow\downarrow}/2\right)}$) when intervalley spin relaxation process dominates, as an example of substrate effects on spin relaxation of strong SOC Dirac materials in Ref.~\citenum{xu2023substrate}. Such angle relates to the spin expectation value direction between initial $\uparrow$ and final $\downarrow$ states. 

\subsection{Dyakonov-Perel (DP) and Free induction decay (FID) mechanisms\label{subsec:DP_FID_mechanism}}

\subsubsection{Model relations}

For non-magnetic materials, with nonzero k-dependent internal magnetic field ${\bf B}_{k}^{\mathrm{in}}$
induced by inversion symmetry broken and spin-orbit coupling, the spins at ${\bf k}$ precess
about ${\bf B}_{k}^{\mathrm{in}}$. The Larmor precession frequency
vector can be defined as:

\begin{align}
{\bf \Omega}_{k}= & \Delta_{k}\widehat{{\bf S}}_{k}^{\mathrm{exp}},\label{eq:Larmor_frequency}
\end{align}
where $\widehat{{\bf S}}_{k}^{\mathrm{exp}}$ is the normalized ${\bf S}_{k}^{\mathrm{exp}}$.

We define ${\bf \Omega}_{\perp\widehat{{\bf r}}}$ as the component of
${\bf \Omega}$ perpendicular to direction $\widehat{{\bf r}}$. Suppose
the fluctuation amplitude among different k-points of ${\bf \Omega}_{\perp\widehat{{\bf r}}}$
is $\Delta{\bf \Omega}_{\perp\widehat{{\bf r}}}$ and numerically
we define it as (using Eq. \ref{eq:thermal_average}) 
\begin{align}
\Delta{\bf \Omega}_{\perp\widehat{{\bf r}}}= & \sqrt{\left\langle \left|{\bf \Omega}_{\perp\widehat{{\bf r}}}-\left\langle {\bf \Omega}_{\perp\widehat{{\bf r}}}\right\rangle \right|^{2}\right\rangle }.\label{eq:fluctuation_Larmor}
\end{align}

According to Refs. \citenum{vzutic2004spintronics} and \citenum{wu2010spin},
a nonzero $\Delta{\bf \Omega}$ leads to finite spin lifetime $\tau_{s}^{\Delta{\bf \Omega}}$
along ${\bf \widehat{r}}$ and the spin relaxation mechanism depends
on the magnitude of $\tau_{p}\Delta{\bf \Omega}$\citep{wu2010spin,vzutic2004spintronics}
(the subindex ``$\perp\widehat{{\bf r}}$'' is dropped for simplicity):

(i) DP mechanism if $\tau_{p}\Delta{\bf \Omega}\ll1$ (strong scattering
limit). We have the DP relation
\begin{align}
\left(\tau_{s}^{\Delta\Omega}\right)^{-1}\sim & \left(\tau_{s}^{\mathrm{DP}}\right)^{-1}\sim\tau_{p}\left(\Delta{\bf \Omega}\right)^{2}.\label{eq:DP}
\end{align}

(ii) FID mechanism if $\tau_{p}\Delta\Omega\gtrsim1$ (weak scattering
limit). We have
\begin{align}
\left(\tau_{s}^{\Delta\Omega}\right)^{-1}\sim & \left(\tau_{s}^{\mathrm{FID}}\right)^{-1}\sim\Delta{\bf \Omega}.\label{eq:free}
\end{align}

(iii) Between (i) and (ii) regimes, there isn't a good approximate
relation for $\left(\tau_{s}^{\Delta\Omega}\right)^{-1}$, but we
may expect that\citep{wu2010spin}
\begin{equation}
\left(\tau_{s}^{\mathrm{DP}}\right)^{-1}<\left(\tau_{s}^{\Delta\Omega}\right)^{-1}<\left(\tau_{s}^{\mathrm{FID}}\right)^{-1}.\label{eq:between_free_DP}
\end{equation}

\subsubsection{Land\'e g-factor of free carriers\label{subsec:g-factor}}

Besides intrinsic spin-orbit coupling, one channel to induce nonzero $\Delta{\bf \Omega}$ is through $g$ factor fluctuations under magnetic field. In Sec.~\ref{subsec:coherent_dynamics_and_external_fields}, we have
described how ${\bf B}^{\mathrm{ext}}$ is considered in FPDM approach,
therefore, with $H_{z}\left({\bf B}^{\mathrm{ext}}\right)$, the magnetic-field
effects on spin dynamics are straightforwardly included in FPDM simulations.

For the description of the magnetic-field effects, Land\'e $g$-factor
has been widely used\citep{xu2023spin}. For a single band and a pair
of bands, $g$-factor is well defined and relates to ${\bf B}^{\mathrm{ext}}$-induced
change of energy, energy splitting and/or Larmor precession frequency.
In this work, we limit our discussions to two Kramers degenerate bands
under a transverse ${\bf B}^{\mathrm{ext}}$ (perpendicular to the
direction of the excess/excited spin ${\bf S}\equiv\left(S_{x},S_{y},S_{z}\right)$).
In general, in the two-band case, $g$-factor is a tensor and may
be defined by the relation ${\bf \Omega}_{k}\left({\bf B}^{\mathrm{ext}}\right)=\mu_{B}{\bf B}^{\mathrm{ext}}g_{k}^{S}$.
Here, for simplicity, we assume ${\bf \Omega}_{k}\left({\bf B}^{\mathrm{ext}}\right)||{\bf B}^{\mathrm{ext}}$,
which is valid in many materials including CsPbBr$_{3}$. With this
assumption and from Eq. \ref{eq:Larmor_frequency},
\begin{align}
g_{k}^{S}= & \frac{{\bf \Omega}_{k}\left({\bf B}^{\mathrm{ext}}\right)\cdot\widehat{{\bf B}}^{\mathrm{ext}}}{\mu_{B}B^{\mathrm{ext}}}.\label{eq:gfacS}
\end{align}

${\bf \Omega}_{k}\left({\bf B}^{\mathrm{ext}}\right)$ is computed
using Eq. \ref{eq:Larmor_frequency}.

However, in many previous theoretical studies\citep{yu2016effective,kirstein2022lande},
$g$-factors were defined based on pseudo-spins related to the total
magnetic momenta $J^{\mathrm{at}}$, which are determined from the
atomic-orbital models. The pseudo-spins can have opposite directions
to the actual spins ($S$). Most previous experimental studies adopted the
same convention for the signs of carrier $g$-factors. Therefore,
to compare with $g$-factors obtained in those previous studies, we
introduce a correction factor $C^{\mathrm{S\rightarrow J}}$ and define
a new $g$-factor:
\begin{align}
\widetilde{g}_{k}\left(\widehat{{\bf B}}^{\mathrm{ext}}\right)= & C^{S\rightarrow J}g_{k}^{S}.\label{eq:gfac}
\end{align}

$C^{\mathrm{S\rightarrow J}}=m_{S}^{\mathrm{at}}/m_{J}^{\mathrm{at}}$
with $m_{J}^{\mathrm{at}}$ and $m_{S}^{\mathrm{at}}$ the total and
spin magnetic momenta respectively, obtained from the atomic-orbital
model\citep{kirstein2022lande}. $C^{\mathrm{S\rightarrow J}}$ is
independent from k-point, and is $\mp$1 for electrons and holes respectively
for CsPbBr$_{3}$~\citep{xu2023spin}.

As $\widetilde{g}_{k}$ is different at different ${\bf k}$, we can
define its fluctuation amplitude as 
\begin{align}
\Delta\widetilde{g}= & \left\langle \left(\widetilde{g}-\left\langle \widetilde{g}\right\rangle \right)^{2}\right\rangle .\label{eq:g_sigma}
\end{align}

From the above equation, we have the fluctuation amplitude of ${\bf \Omega}_{k}\left({\bf B}^{\mathrm{ext}}\right)$
\begin{align}
\Delta\Omega\left({\bf B}^{\mathrm{ext}}\right)= & \mu_{B}B^{\mathrm{ext}}\Delta\widetilde{g}.\label{eq:dg_induced_dOmega}
\end{align}

A nonzero $\Delta\Omega\left({\bf B}^{\mathrm{ext}}\right)$ leads
to spin dephasing under external magnetic field ${\bf B}^{\mathrm{ext}}$ and the mechanism may be DP or FID depending on
the magnitude of $\tau_{p}\Delta{\bf \Omega}$ as discussed above.

\subsection{Determination of spin relaxation mechanism}

In general, the applicability of our FPDM approach does not depend on specific spin relaxation mechanism. To determine the dominant relaxation or dephasing mechanism, we have the following two approaches by utilizing our FPDM calculations. 

\subsubsection{Method 1: Comparing FPDM calculations with FGR or model relations}

The first approach is to compare FPDM results directly with simple models designed for various mechanisms as introduced in previous two sections. If FPDM agrees with one of the model relations, it is a good indicator of dominant mechanism. 

\paragraph{1.1 EY mechanism.}

Since spin precession is suppressed when EY mechanism dominates, the
DM master equation in semiclassical limit from Eq. \ref{eq:semiclassical}
and FGR formula with spin-flip scattering Eq. \ref{eq:FGR-1} should
describe well spin relaxation (suppose spin matrix is diagonalized
in degenerate subspaces along the direction of the excess spin). Therefore,
if the values and trends of $\tau_{s}$ by Eq. \ref{eq:semiclassical}
and \ref{eq:FGR-1} are similar to those obtained from FPDM calculations, the dominating mechanism is
likely the EY.

\paragraph{1.2 DP or FID mechanism.}

With nonzero $\Delta{\bf \Omega}$, whether spin relaxation is dominated
by DP or FID mechanism may be determined by comparing the values and
trends of $\tau_{s}$ by Eq. \ref{eq:DP} or \ref{eq:free} with
the FPDM results. We have shown the success of such analysis in our previous work on spin relaxation and transport in germanene and silicene under electric field~\citep{xu2021giant} as well as CsPbBr$_3$ under magnetic field~\citep{xu2023spin}.

\subsubsection{Method 2: Tuning the scattering strength or precession frequency in FPDM calculations}

Conventionally, spin relaxation mechanism is determined from the relation
between $\tau_{s}$ and $\tau_{p}$ - EY mechanism leads to $\tau_{s}\propto\tau_{p}$
while DP mechanism leads to $\tau_{s}\propto\tau_{p}^{-1}$. Another way to look at this is through tuning the strength of scatterings (e.g. physically, increasing impurity concentration increases scattering). If $\tau_{s}$ decreases with increasing scattering, it's likely EY; otherwise, it's likely DP mechanism, because $\tau_{p}$ always decreases with increasing scatterings. 

Practically in our FPDM calculations, this approach can be implemented by introducing a scaling factor $F^{\mathrm{sc}}$ 
to tune the scattering strength, i.e.
multiplying $F^{\mathrm{sc}}$ to the scattering term $\frac{d\rho}{dt}|_{\mathrm{scatt}}$
of the DM master equation (Eq. \ref{eq:dynamics}) but keep the coherent
term (the first term of Eq. \ref{eq:dynamics}) unchanged. This is
equivalent to multiply $F^{\mathrm{sc}}$ to all elements of the generalized
scattering-rate matrix $P$ (Eq. \ref{eq:Peph}, \ref{eq:Pei} and
\ref{eq:Pee}). 

To avoid confusions , we name carrier and spin lifetimes after introducing
$F^{\mathrm{sc}}$ as $\tau_{s}'$ and $\tau_{p}'$ respectively.
According to Eq. \ref{eq:carrier_lifetime},
we always have $\tau_{p}'/\tau_{p}=\left(F^{\mathrm{sc}}\right)^{-1}$. 
On the other hand, $\tau_{s}'$ depends on the spin relaxation mechanism. 
For EY mechanism, $\tau_{s}'/\tau_{p}'=\tau_{s}/\tau_{p}=C^{\mathrm{EY}}$, where $C^{\mathrm{EY}}$ is a constant unrelated to $F^{\mathrm{sc}}$.
Therefore, it can be proven that $\tau_{s}'/\tau_{s}=\left(F^{\mathrm{sc}}\right)^{-1}=\tau_{p}'/\tau_{p}$.
Similarly, for DP mechanism, we can prove that $\tau_{s}'/\tau_{s}=F^{\mathrm{sc}}=\left(\tau_{p}'/\tau_{p}\right)^{-1}$.
For FID mechanism, as $\tau_{s}'$ is irrelevant to $\tau_{p}'$,
we have $\tau_{s}'/\tau_{s}\equiv1$. Therefore, the relation between
$\tau_{s}'/\tau_{s}$ and $\tau_{p}'/\tau_{p}$ by tuning $F^{\mathrm{sc}}$ is useful to understand
spin relaxation mechanism.

Moreover, the relation between $\tau_{s}$ and $\Delta\Omega$ is
also useful to understand spin relaxation mechanism. $\Delta\Omega$
can be tuned easily by tuning the energy splitting.

\section{Applications}

\subsection{Non-magnetic materials with inversion symmetry}

\begin{figure*}
\includegraphics[scale=0.71]{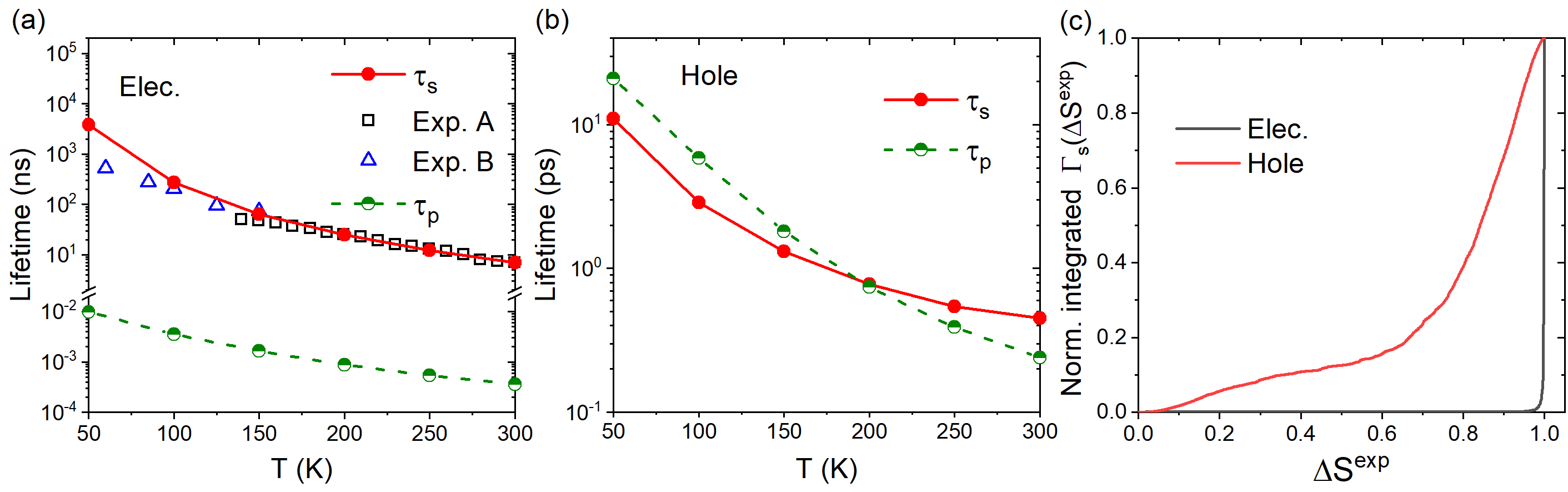}

\caption{Spin relaxation in Silicon (Si). Spin ($\tau_{s}$) and carrier ($\tau_{p}$)
lifetimes of (a) electrons and (b) holes of Si. Exp. A and B are from
Ref. \citenum{lepine1972spin} and \citenum{huang2007coherent}. (c)
Cumulative contributions to spin relaxation by change in spin, $\Delta$s,
per scattering event defined based on Eqs.~\ref{eq:GenFGR-1}-\ref{eq:GenFGR-3}: electrons in Si exhibit
spin flips with all contributions at $\Delta$s = 1, whereas holes
in Si and electrons in iron exhibit a broad distribution in $\Delta$s~\citep{xu2020spin}.\label{fig:silicon}}
\end{figure*}

We first present results for systems with inversion symmetry traditionally
described by EY mechanism. Figure~\ref{fig:silicon}(a) shows that
our theoretical electron $\tau_{s}$ of Si as a function of $T$ are
in excellent agreement with experimental measurements\citep{lepine1972spin,huang2007coherent}.
Note that previous first-principles calculations\citep{restrepo2012full}
approximated spin-flip electron-phonon matrix elements from pseudospin
wavefunction overlap and spin-conserving electron-phonon matrix element,
effectively assuming that the scattering potential varies slowly on
the scale of a unit cell; we make no such approximation in our 
FPDM approach. 
In contrast, holes in Si exhibit strong spin mixing with spin-2/3
character and spin expectation values no longer close to $\hbar/2$.
Figure~\ref{fig:silicon}(b) shows our predictions for the hole spin
lifetime $\tau_{s,h}$ which is much shorter than the electron spin
as a result of the strong mixing (450~fs for holes compared to 7~ns
for electrons at 300~K) and is much closer to carrier relaxation time $\tau_{p}$. Additionally,
Fig.~\ref{fig:silicon}(c) shows that the change in spin expectation
values ($\Delta S^{\mathrm{exp}}$) per scattering event (evaluated
using Eqs.~\ref{eq:GenFGR-1} -\ref{eq:GenFGR-3}) has a broad distribution
for holes in Si, indicating that they cannot be described purely by
spin-flip transitions, while conduction electrons in Si predominantly
exhibit spin-flip transitions with $\Delta S^{\mathrm{exp}}=1$.

\subsection{Materials with high spin polarization}

In ferromagnets and antiferromagnets, spins are highly polarized,
so that spin relaxation is likely dominated by EY mechanism. In Ref. \citenum{xu2020spin},
we simulated $\tau_{s}$ of bcc iron, where we found 
good agreement with experiments, with a dominant EY mechanism (magnon was not considered here). In nonmagnetic materials, when inversion
symmetry is broken and SOC strength is moderate or large, it
has been found that some of them such as transition metal dichalcogenides  also have highly polarized
carrier spins and exhibit EY spin relaxation.

\subsubsection{Transition metal dichalcogenides (TMDs)}

Monolayer TMDs exhibit exciting features including spin-polarized
bands, valley-specific optical selection rules and spin-valley locking.
In Ref. \citenum{dey2017gate,li2021valley}, it has been shown that
by introducing doping in monolayer TMDs, ultraslow decays of Kerr
rotations, which correspond to ultralong spin/valley lifetimes of
resident carriers especially resident holes, can be observed at low temperatures.
Those features establish TMDs' advantages for spin-valleytronics
and (quantum) information processing. Although spin/valley relaxation time
of resident carriers in monolayer TMDs, most relevant to
spinvalleytronics' application, has been extensively
examined, the underlying relaxation mechanisms especially the effects
of different types of impurities have not been addressed through 
\emph{ab-initio} simulations.

In Ref. \citenum{xu2021ab}, we clarified the above problem by conducting
\emph{ab initio} real-time dynamics simulations with relevant scattering
mechanisms. We focused on WSe$_{2}$ due to its larger valence band
SOC splitting and weaker interlayer coupling compared with other TMDs
and focused on dynamics of resident holes as $\tau_{s}$ of holes
seem longer than electrons.

For holes of monolayer WSe$_{2}$, spin/valley relaxation is completely
determined by intervalley spin-flip scattering between $K$ and $K'$
valleys because of spin-valley locking. Previously, we reported spin/valley
lifetimes of resident holes of monolayer TMDs at $T\geq$50 K with
e-ph scattering\citep{xu2020spin}. At very low temperatures, e.g.,
10 K, intervalley e-ph scattering is however not activated as the
corresponding phonon occupation is negligible, so that other scattering
mechanisms need to be included. Note that e-e scattering should not play
an important role in spin relaxation of holes of TMDs. The reason
is: The e-e scattering is a two-particle process where a transition
is accompanied by another transition with energy and momentum being
conserved. Considering the fact that only the highest occupied band
is involved in dynamics of TMD holes, for a e-e process, a $K$$\rightarrow$$K'$
($K'$$\rightarrow$$K$) spin-flip transition must be accompanied
by an opposite $K'$$\rightarrow$$K$ ($K$$\rightarrow$$K'$) spin-flip
transition. Overall, an e-e scattering process cannot change the spin
of the system. Therefore, we will include only e-ph and e-i scatterings.

\begin{figure}
\includegraphics[scale=0.15]{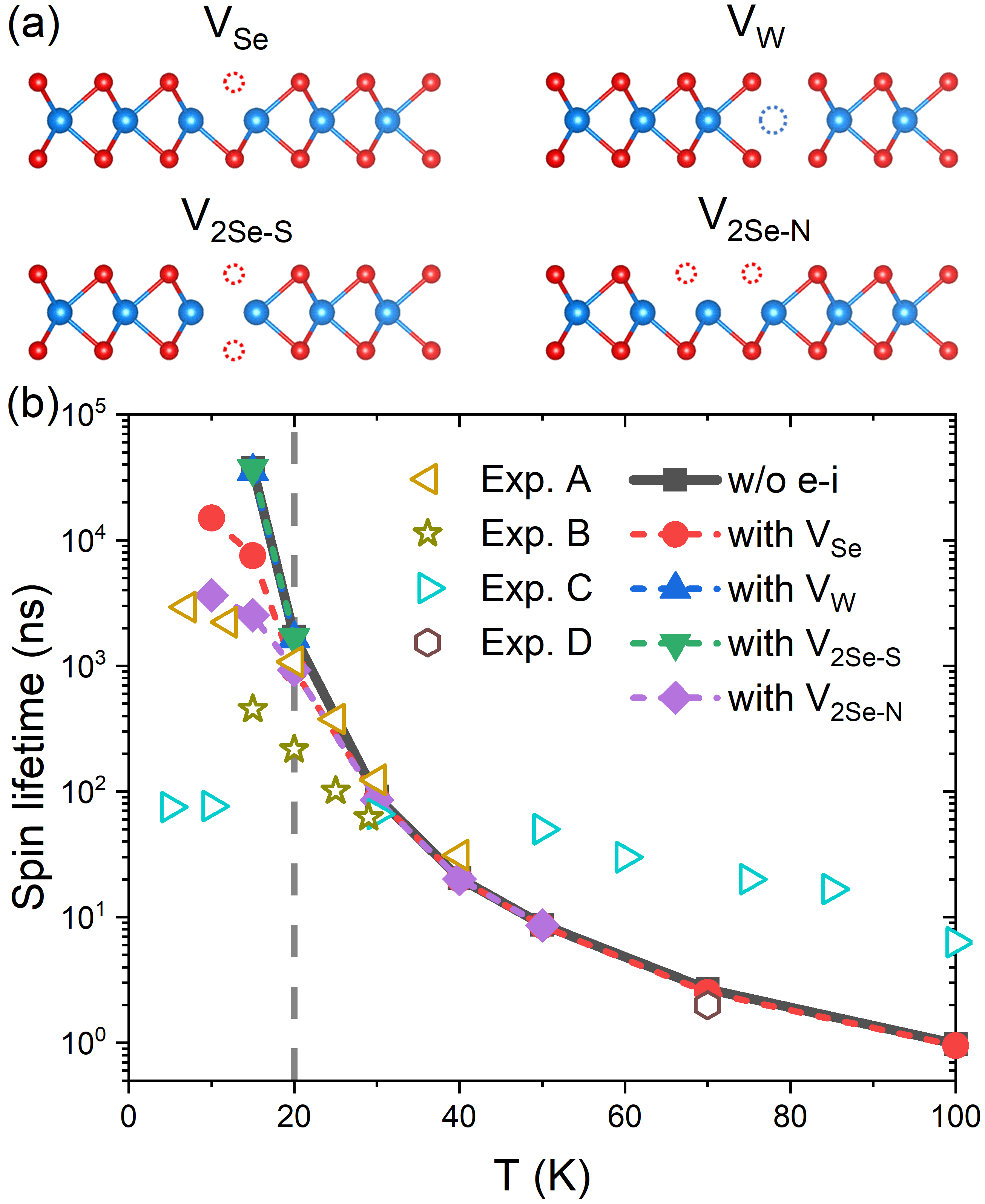}

\caption{(a) The schematics of four types of impurities in WSe$_{2}$. (b)
Spin lifetimes of holes of monolayer WSe$_{2}$ with a relatively
low hole density $10^{11}$ cm$^{-2}$ with impurities compared with
experimental data\citep{li2021valley,goryca2019detection,song2016long,yan2017long}.
The choices of impurity concentration $n_{i}$ of different impurities
are given in the main text~\citep{xu2021ab}.\label{fig:ml-wse2}}
\end{figure}

Experimentally there exists many types of impurities/defects in TMD
samples. Here we pick four types of impurities with different symmetries
and chemical bonds (see Fig. \ref{fig:ml-wse2}(a)) - Se vacancy (Se
vac.), two neighboring Se vacancies (2N-Se vac.), W vacancy (W vac.)
and two Se vacancies with the same in-plane position (2S-Se vac.).
According to Refs. \citenum{edelberg2019approaching,rhodes2019disorder,yankowitz2015local},
$n_{i}$ ranges from 8$\times$10$^{10}$ to 10$^{14}$ cm$^{-2}$
depending on samples. Considering that Se vac. is often regarded as
the most abundant impurity, we choose a reasonable $n_{i}$ of Se
vac. - 7$\times$10$^{11}$ cm$^{-2}$, which is taken relatively
low for better comparison with experimental $\tau_{s}$ shown in Fig.
\ref{fig:ml-wse2}(b). Since the formation energy of Se vac. seems
lowest and $n_{i}$ of larger impurities are found lower than smaller
impurities\citep{yankowitz2015local}, $n_{i}$ of 2N-Se vac. is chosen
as 8$\times$10$^{9}$ cm$^{-2}$ lower than Se vac. and also for
better comparison with experimental $\tau_{s}$. $n_{i}$ of W vac.
and 2S-Se vac. are chosen arbitrarily as we found they have rather
weak effects on spin relaxation and are 7$\times$10$^{11}$ and 3.5$\times$10$^{11}$
cm$^{-2}$, respectively.

From Fig. \ref{fig:ml-wse2}, we first find that assuming $n_{i}$
not so high, at $T$\textgreater 20 K, spin relaxation is almost
driven by e-ph scattering and impurities can only affect spin relaxation
at $T$$\leq$20 K. For the effects on spin relaxation of different
impurities, we have 2N-Se vac. $\gg$ Se vac. $\gg$ W vac. $\sim$
2S-Se vac.. Moreover, the temperature dependence of $\tau_{s}$ with
2N-Se vac. is much weaker and in better agreement with experiments
than that with Se vac.. Therefore, the observed weak temperature dependence
in some experiments is probably related the existence of larger impurities
with lower symmetries. Our observations suggest that the local symmetry
and chemical bonds surround a impurity have large impact on spin relaxation.
To understand why the effects of different impurities on spin relaxation
significantly differ, further theoretical investigations on how impurities
affect impurity potentials and their SOC corrections are required.

\subsubsection{Germanene under an electric field or on a substrate\label{subsec:ge-substrates}}

In Ref. \citenum{xu2021giant}, through FPDM simulations, we predicted
that monolayer germanene (ML-Ge) has giant spin lifetime anisotropy,
spin-valley-locking (SVL) effect under  nonzero perpendicular electric
field $E_{z}$ and long $\tau_{s}$ ($\sim$100 ns at 50 K), which
makes it advantageous for spin-valleytronic applications. In 2D-material-based
spintronic devices, the materials are usually supported on a substrate.
Therefore, for the design of those devices, it is crucial to understand
substrate effects on spin relaxation. We further examined $\tau_{s}$
of ML-Ge on different substrates in Ref. \citenum{xu2023substrate}.

\begin{figure*}[!t]
\includegraphics[scale=0.54]{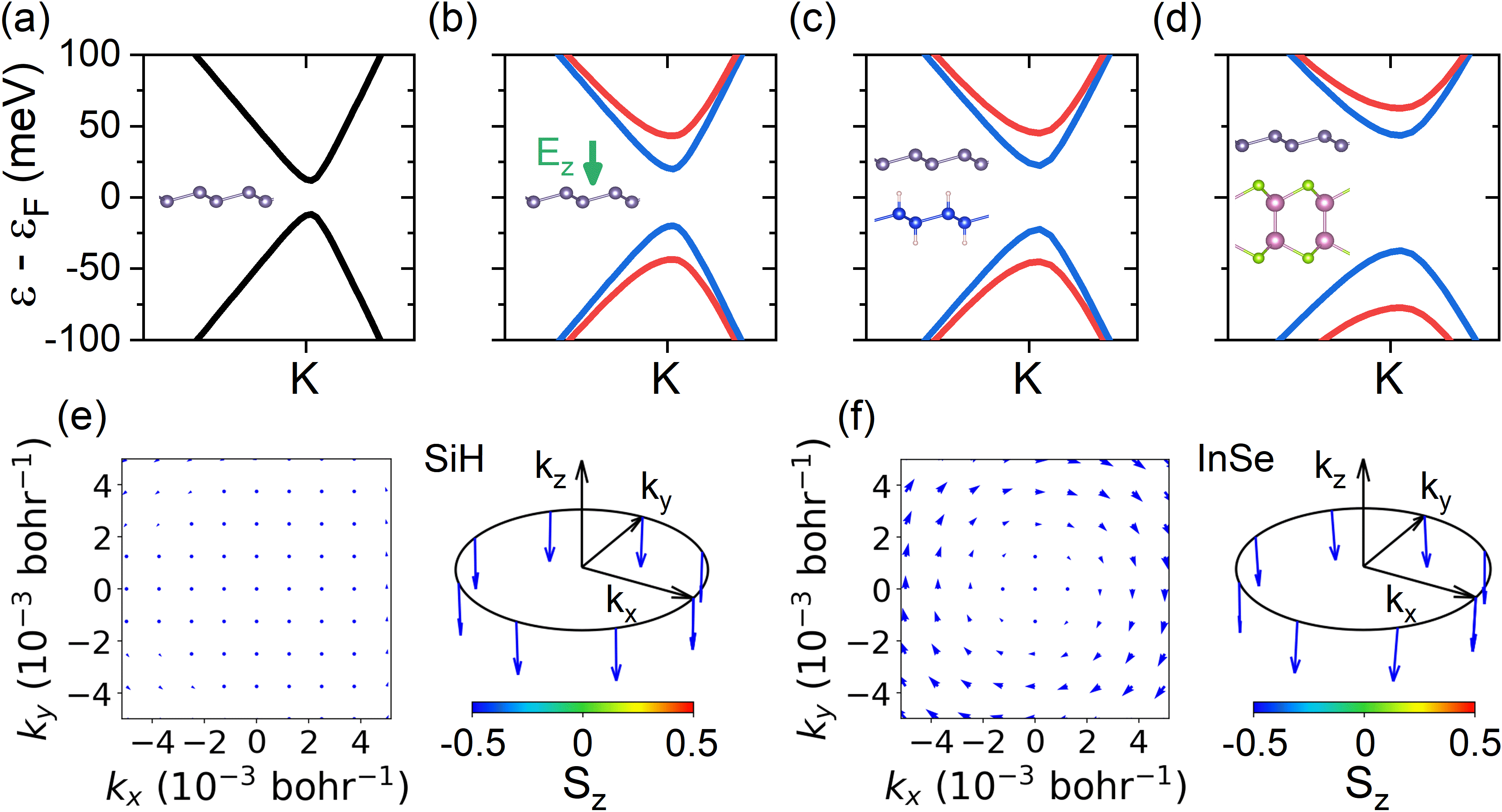}

\caption{Band structures and spin textures around the Dirac cones of ML-Ge
systems with and without substrates. (a)-(d) show band structures
of ML-Ge under $E_{z}=0$ and under -7 V/nm and ML-Ge on silicane
(SiH) and on InSe substrates respectively. The red and blue bands
correspond to spin-up and spin-down states. Due to time-reversal symmetry,
band structures around another Dirac cone at $K'=-K$ are the same
except that the spin-up and spin-down bands are reversed. The grey,
white, blue, pink and green balls correspond to Ge, H, Si, In and
Se atoms, respectively. Band structures of ML-Ge on germanane (GeH)
and GaTe are shown in Supplementary Figure 4 in the Supporting Information,
and are similar to those of ML-Ge on SiH and InSe substrates, respectively.\textcolor{blue}{{}
}(e) and (f) show spin textures in the $k_{x}$-$k_{y}$ plane and
3D plots of the spin vectors ${\bf S}_{k_{1}}^{\mathrm{exp}}$ on
the circle $|\protect\overrightarrow{k}|=0.005$ bohr$^{-1}$ of the
band at the band edge around $K$ of ML-Ge on SiH and InSe substrates
respectively. The color scales $S_{z}^{\mathrm{exp}}$ and the arrow
length scales the vector length of in-plane spin expectation value~\citep{xu2021giant}.\label{fig:ml-ge_bands}}
\end{figure*}

We show band structures and spin textures of free-standing and supported
ML-Ge in Fig.~\ref{fig:ml-ge_bands}, which are essential for understanding
spin relaxation mechanisms. At $E_{z}=0$, ML-Ge has time-reversal
and inversion symmetries, so that its bands are Kramers degenerate\citep{vzutic2004spintronics}.
A finite $E_{z}$ or a substrate breaks the inversion symmetry and
induces a strong out-of-plane ${\bf B}^{\mathrm{in}}$ (and also ${\bf S}^{\mathrm{exp}}$,
Eq.~\ref{eq:Bin}), which splits the Kramers pairs into spin-up and
spin-down bands\citep{xu2021giant}. Interestingly, we find that band
structures of ML-Ge-SiH (Fig.~\ref{fig:ml-ge_bands}c) and ML-Ge-GeH
(not shown) are quite similar to free-standing ML-Ge under $E_{z}$=-7
V/nm (ML-Ge@-7V/nm, Fig.~\ref{fig:ml-ge_bands}b), which indicates
that the impact of the SiH/GeH substrate on band structure (and also
${\bf B}^{\mathrm{in}}$) may be similar to a finite $E_{z}$. This
similarity is frequently assumed in model Hamiltonian studies\citep{van2016spin,ertler2009electron}.
On the other hand, the band structures of ML-Ge-InSe (Fig.~\ref{fig:ml-ge_bands}d)
and ML-Ge-GaTe (not shown) have more differences from the free-standing
one under $E_{z}$, with larger band gaps, smaller band curvatures
at Dirac Cones, and larger electron-hole asymmetry of band splittings.
This implies that the impact of the InSe/GaTe substrates can not be
approximated by applying an $E_{z}$ to the free-standing ML-Ge, unlike
SiH/GeH substrates. We further examine ${\bf S}^{\mathrm{exp}}$ of
substrate-supported ML-Ge. Importantly, from Fig.~\ref{fig:ml-ge_bands}e
and \ref{fig:ml-ge_bands}f, although ${\bf S}^{\mathrm{exp}}$ of
ML-Ge on substrates are highly polarized along $z$ (out-of-plane)
direction, the in-plane components of ${\bf S}^{\mathrm{exp}}$ of
ML-Ge-InSe (and ML-Ge-GaTe) are much more pronounced than ML-Ge-SiH
(and ML-Ge-GeH). Such differences are crucial to the out-of-plane
spin relaxation as discussed later.

\begin{figure}[!t]
\includegraphics[scale=0.32]{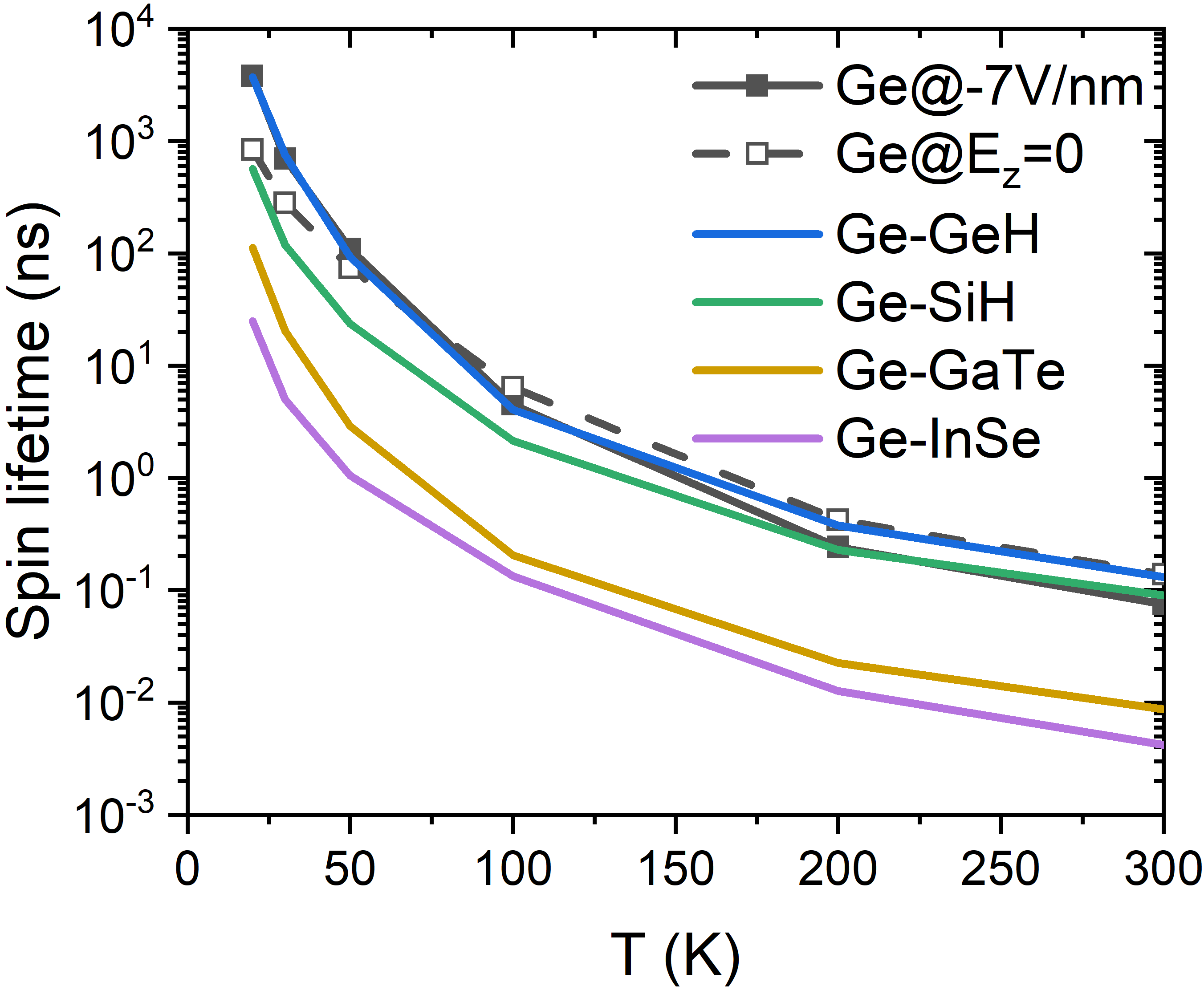}

\caption{The out-of-plane $\tau_{s}$ of ML-Ge under $E_{z}=0$, -7 V/nm and
substrate-supported ML-Ge as a function of temperature without impurities.
Here we show electron $\tau_{s}$ for intrinsic ML-Ge systems except
that hole $\tau_{s}$ is shown for ML-Ge-InSe, since electron $\tau_{s}$
are longer than hole $\tau_{s}$ at low $T$ except ML-Ge-InSe~\citep{xu2023substrate}.\label{fig:T1_ge}}
\end{figure}

We compare out-of-plane $\tau_{s}$ due to e-ph scattering between
the free-standing ML-Ge (with/without an electric field) and ML-Ge
on different substrates in Fig.~\ref{fig:T1_ge}. From Fig.~\ref{fig:T1_ge},
we find that $\tau_{s}$ of ML-Ge under $E_{z}=0$ and -7 V/nm are
at the same order of magnitude for a wide range of temperatures. The
differences are only considerable at low $T$, e.g, by 3-4 times at
20 K. On the other hand, $\tau_{s}$ of supported ML-Ge are very sensitive
to the specific substrates. While $\tau_{s}$ of ML-Ge-GeH and ML-Ge-SiH
have the same order of magnitude as the free-standing ML-Ge, in particular
very close between ML-Ge-GeH and ML-Ge@-7 V/nm, $\tau_{s}$ of ML-Ge-GaTe
and ML-Ge-InSe are shorter by at least 1-2 orders of magnitude in
the whole temperature range.

Since spins of ML-Ge at $E_{z}\neq0$ and on a substrate are highly
polarized, spin relaxation in ML-Ge systems is dominated by EY mechanism
caused by spin-flip scattering. According to Eq. \ref{eq:taus_approx}
and \ref{eq:taus_b2_relation}, $\tau_{s}^{-1}$ is roughly proportional
to density of states (DOS) and spin-mixing parameter $\left\langle b_{z}^{2}\right\rangle $.
Indeed, in Ref. \citenum{xu2023substrate}, we find that at 300 K,
the differences of $\tau_{s}$ of ML-Ge on different substrates are
well explained by the differences of the products of DOS and $\left\langle b_{z}^{2}\right\rangle $.
However, at $T\le$50 K, the differences of the products of DOS and
$\left\langle b_{z}^{2}\right\rangle $ for different substrates are
about 3-7 times, while the differences of $\tau_{s}$ for different
substrates can be as large as two orders of magnitude. Therefore,
the substrate effects on $\tau_{s}$ can not be fully explained by
the changes of DOS and $\left\langle b_{z}^{2}\right\rangle $, in
particular at relatively low $T$.

\begin{figure}[!t]
\includegraphics[scale=0.35]{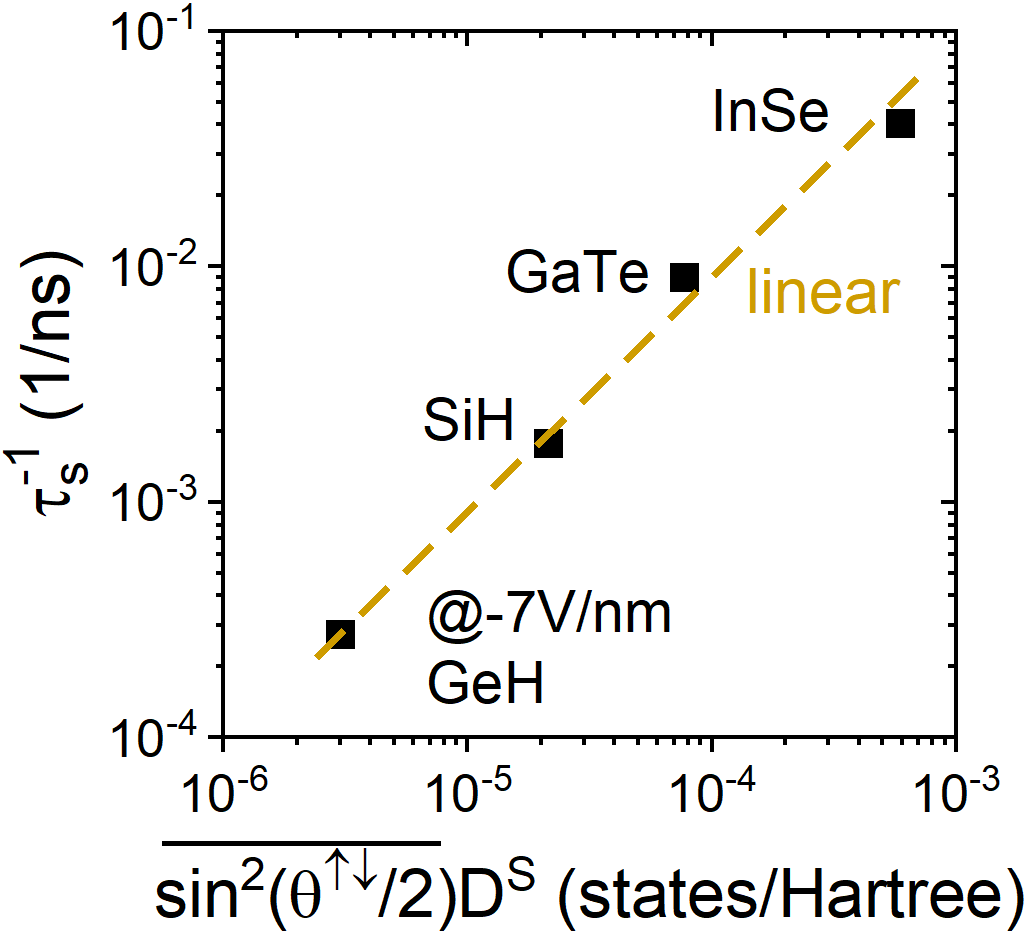}

\caption{The relation between $\tau_{s}^{-1}$ and $\overline{\mathrm{sin}^{2}\left(\theta^{\uparrow\downarrow}/2\right)}$
multiplied by the scattering density of states $D^{S}$ at 20 K. $D^{S}$
is defined in \ref{eq:DS}. $\theta^{\uparrow\downarrow}$ is the
spin-flip angle between two electronic states. For two states $\left(k,n\right)$
and $\left(k',n'\right)$ with opposite spin directions, $\theta^{\uparrow\downarrow}$
is the angle between $-{\bf S}_{kn}^{\mathrm{exp}}$ and ${\bf S}_{k'n'}^{\mathrm{exp}}$.
$\overline{\mathrm{sin}^{2}\left(\theta^{\uparrow\downarrow}/2\right)}$
is defined in Eq. \ref{eq:sin2}. The variation of $D^{S}$ among
different substrates is at most three times, much weaker than the
variations of $\tau_{s}^{-1}$ and other quantities shown here~\citep{xu2023substrate}.\label{fig:spin-flip_matrix_elements}}
\end{figure}

Since $\tau_{s}^{-1}\propto\overline{|\widetilde{g}^{\uparrow\downarrow}|^{2}}D^{\mathrm{S}}$
(Eq. \ref{eq:taus_approx}), to understand substrate effects on $\tau_{s}$
at low $T$, we first compare $\tau_{s}^{-1}$ and $\overline{|\widetilde{g}^{\uparrow\downarrow}|^{2}}D^{\mathrm{S}}$
of different ML-Ge systems. In Ref. \citenum{xu2023substrate}, we
found that $\tau_{s}^{-1}$ is almost linearly proportional to $\overline{|\widetilde{g}^{\uparrow\downarrow}|^{2}}D^{S}$
at 20 K. As the variation of $D^{S}$ among ML-Ge on different substrates
is at most three times, which is much weaker than the large variation
of $\tau_{s}^{-1}$, this indicates that the substrate-induced change
of $\tau_{s}$ is mostly due to the substrate-induced change of spin-flip
matrix elements.

To have deeper intuitive understanding, we then propose an important
electronic quantity for intervalley spin-flip scattering - the spin-flip
angle $\theta^{\uparrow\downarrow}$ between two electronic states.
For two states $\left(k_{1},n_{1}\right)$ and $\left(k_{2},n_{2}\right)$
with opposite spin directions, $\theta^{\uparrow\downarrow}$ is the
angle between $-{\bf S}_{k_{1}n_{1}}^{\mathrm{exp}}$ and ${\bf S}_{k_{2}n_{2}}^{\mathrm{exp}}$
or equivalently the angle between $-{\bf B}_{k_{1}}^{\mathrm{in}}$
and ${\bf B}_{k_{2}}^{\mathrm{in}}$. At low $T$, due to large SOC
splittings of conduction and valence bands of supported ML-Ge, intravalley
spin-flip e-ph scattering processes are forbidden, because the corresponding
phonons have too high energies and are not occupied. So spin relaxation
in supported ML-Ge is dominated by intervalley spin-flip e-ph scattering.
Therefore, $\theta^{\uparrow\downarrow}$ is helpful for understanding
spin relaxation in supported ML-Ge.

The motivation of examining $\theta^{\uparrow\downarrow}$ is that:
Suppose two wavevectors ${\bf k}_{1}$ and ${\bf k}_{2}={\bf -k}_{1}$
are in two opposite valleys $Q$ and -$Q$ respectively and there
is a pair of bands, which are originally Kramers degenerate but splitted
by ${\bf B}^{\mathrm{in}}$. Due to time-reversal symmetry, we have
${\bf B}_{k_{1}}^{\mathrm{in}}=-{\bf B}_{k_{2}}^{\mathrm{in}}$, which
means the two states at the same band $n$ at ${\bf k_{1}}$ and ${\bf k_{2}}$
have opposite spins and $\theta^{\uparrow\downarrow}$ between them
is zero. Therefore, the matrix element of operator $\widehat{A}$
between states $\left(k_{1},n\right)$ and $\left(k_{2},n\right)$
- $A_{k_{1}n,k_{2}n}$ is a spin-flip one and we name it as $A_{k_{1}k_{2}}^{\uparrow\downarrow}$.
According to Ref. \citenum{yafet1963g}, with time-reversal symmetry,
$A_{k_{1}k_{2}}^{\uparrow\downarrow}$ is exactly zero. In general,
for another wavevector ${\bf k_{3}}$ within valley -$Q$ but not
${\bf -k}_{1}$, $A_{k_{1}k_{3}}^{\uparrow\downarrow}$ is usually
non-zero. One critical quantity that determines the intervalley spin-flip
matrix element $A_{k_{1}k_{3}}^{\uparrow\downarrow}$ for a band within
the pair introduced above is $\theta_{k_{1}k_{3}}^{\uparrow\downarrow}$.
Based on time-independent perturbation theory, we can prove that $\left|A^{\uparrow\downarrow}\right|$
between two states is approximately proportional to $\left|\sin\left(\theta^{\uparrow\downarrow}/2\right)\right|$.
The derivation is given in Sec. \ref{subsec:Spin-flip-angle}.

As shown in Figure~\ref{fig:spin-flip_matrix_elements}c, $\tau_{s}^{-1}$
of ML-Ge on different substrates at 20 K is almost linearly proportional
to $\overline{\sin^{2}\left(\theta^{\uparrow\downarrow}/2\right)}D^{S}$,
where $\overline{\sin^{2}\left(\theta^{\uparrow\downarrow}/2\right)}$
is the statistically-averaged modulus square of $\sin\left(\theta^{\uparrow\downarrow}/2\right)$.
This indicates that the relation $\overline{|\widetilde{g}^{\uparrow\downarrow}|^{2}}\propto\overline{\sin^{2}\left(\theta^{\uparrow\downarrow}/2\right)}$
is nearly perfectly satisfied at low $T$, where intervalley processes
dominate spin relaxation. We additionally examined the relation between
$\tau_{s}^{-1}$ and $\overline{\sin^{2}\left(\theta^{\uparrow\downarrow}/2\right)}D^{S}$
at 300 K and found that the trend of $\tau_{s}^{-1}$ is still approximately
captured by the trend of $\overline{\sin^{2}\left(\theta^{\uparrow\downarrow}/2\right)}D^{S}$.

Since $\theta^{\uparrow\downarrow}$ is defined by ${\bf S}^{\mathrm{exp}}$
at different states, $\tau_{s}$ is highly correlated with ${\bf S}^{\mathrm{exp}}$
and more specifically with the anisotropy of ${\bf S}^{\mathrm{exp}}$
(equivalent to the anisotropy of ${\bf B}^{\mathrm{in}}$). Qualitatively,
the larger anisotropy of ${\bf S}^{\mathrm{exp}}$ leads to smaller
$\theta^{\uparrow\downarrow}$ (consistent with Fig. \ref{fig:ml-ge_bands}e
and f) and longer $\tau_{s}$ along the high-spin-polarization direction.
This finding may be applicable to spin relaxation in other materials
whenever intervalley spin-flip scattering dominates or spin-valley
locking exists, e.g., in TMDs\citep{dey2017gate}, Stanene\citep{tao2019two},
2D hybrid perovskites with persistent spin helix\citep{zhang2022room},
etc.

\subsection{DP (Dyakonov-Perel) systems}

In many non-magnetic materials with broken inversion symmetry, spin
relaxation is dominated by DP mechanism.

\subsubsection{GaAs}

Spin dynamics in GaAs has broad interest in spintronics over past
decades\citep{vzutic2004spintronics,kikkawa1998resonant,hilton2002optical,jiang2009electron,kamra2011role}
and more recently\citep{PhysRevX.7.031010,PhysRevLett.121.033902,belykh2018quantum},
partly due to its long spin lifetime in n-type GaAs at low temperatures\citep{kikkawa1998resonant}.
Despite various experimental~\citep{kikkawa1998resonant,hilton2002optical,ohno1999spin,kimel2001room,hohage2006coherent}
and theoretical~\citep{vzutic2004spintronics,yu2005spin,jiang2009electron,mower2011dyakonov,kamra2011role,marchetti2014spin}
(mostly using parameterized model Hamiltonian) studies previously,
the dominant spin relaxation mechanism in bulk GaAs at various $T$
and doping concentrations $n_{i}$ remains unclear.

Through FPDM simulations of $n$-GaAs at various $T$ and $n_{i}$
with different scattering mechanisms in Ref. \citenum{xu2021ab},
we pointed out that although at low temperatures and moderate doping
concentrations e-i scattering dominates carrier relaxation, e-e scattering
is the most dominant process in spin relaxation. Moreover, we found
that the relative contributions of phonon modes vary considerably
between spin and carrier relaxation.

\begin{figure*}
\includegraphics[scale=0.24]{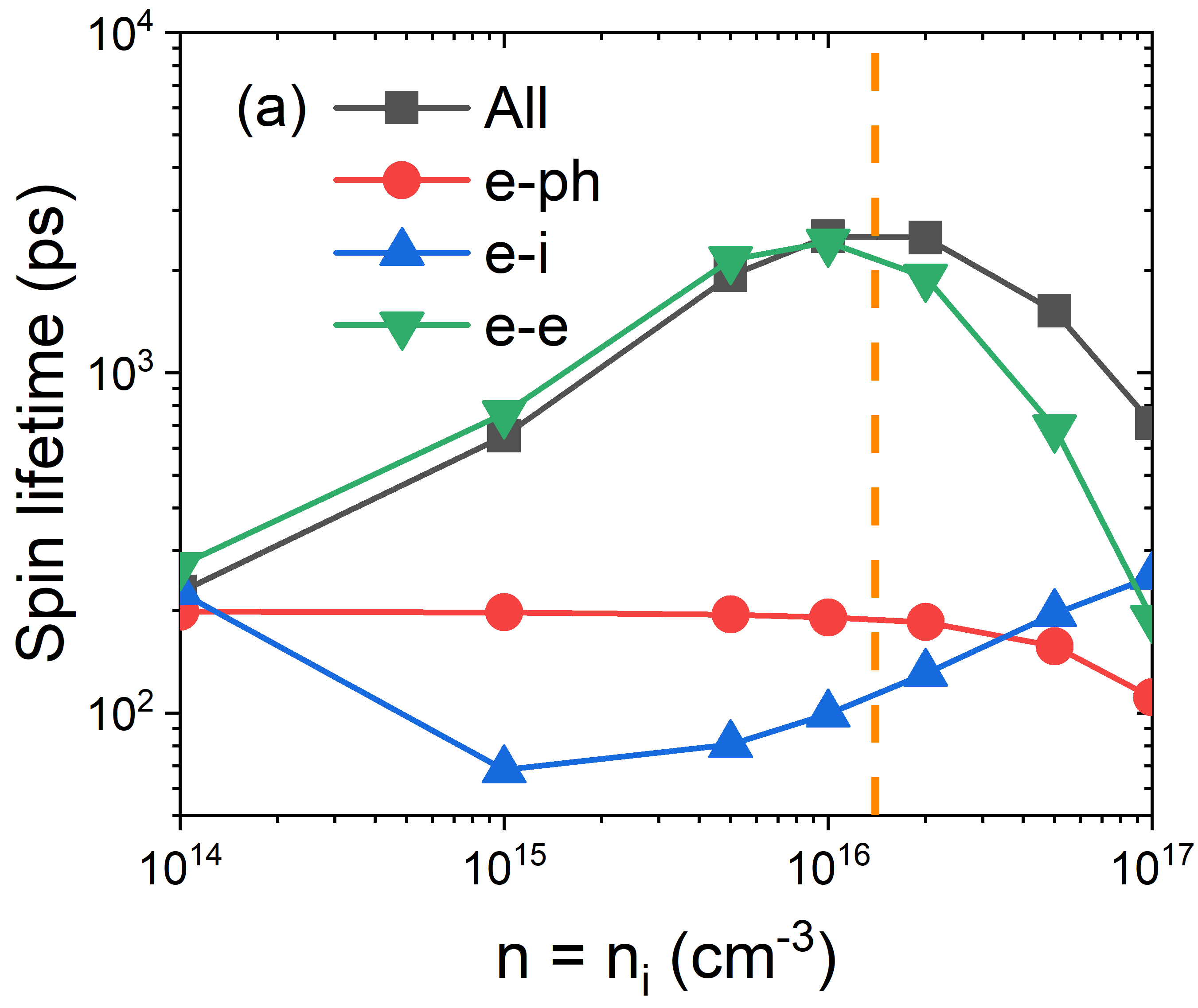} \includegraphics[scale=0.24]{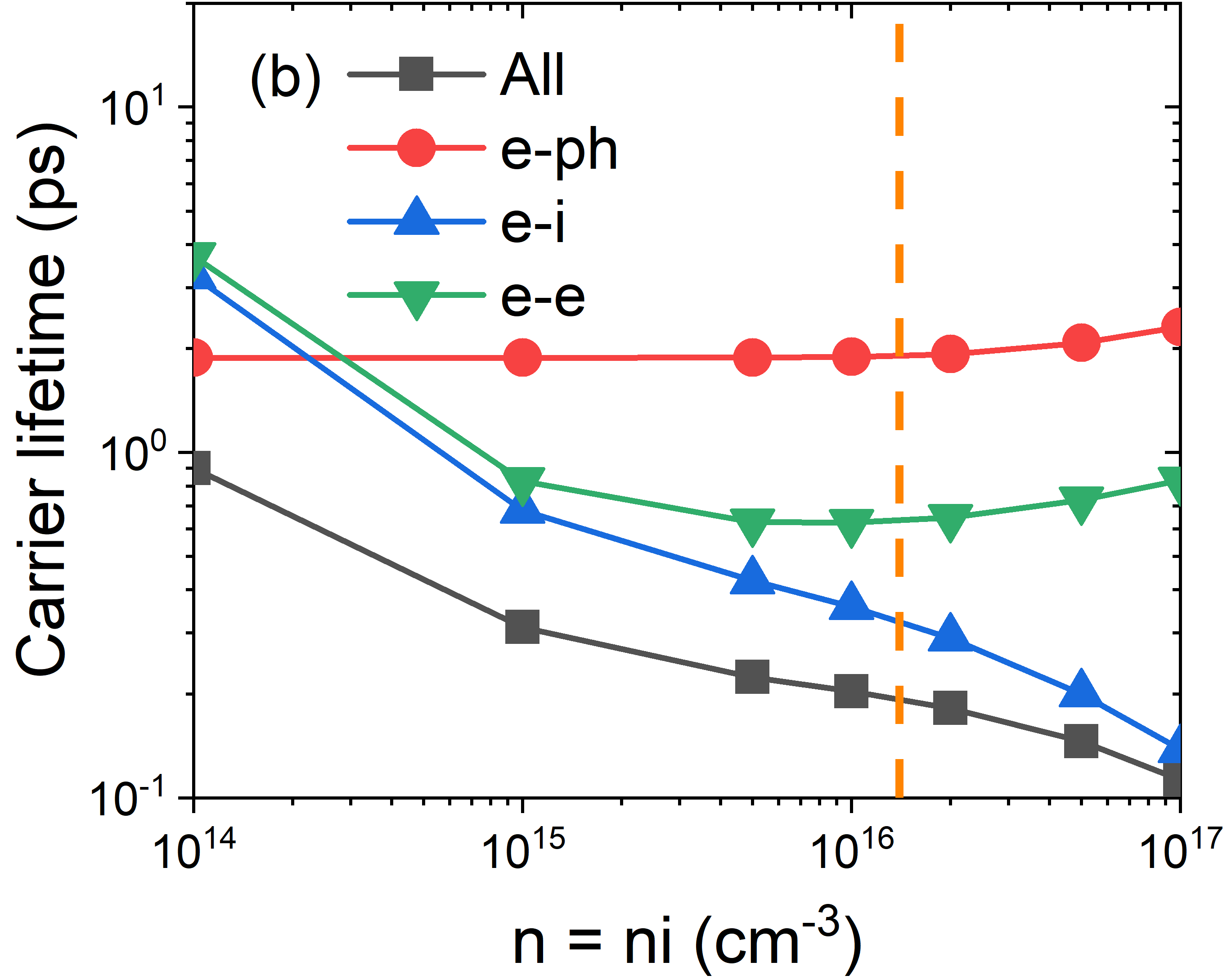}
\includegraphics[scale=0.24]{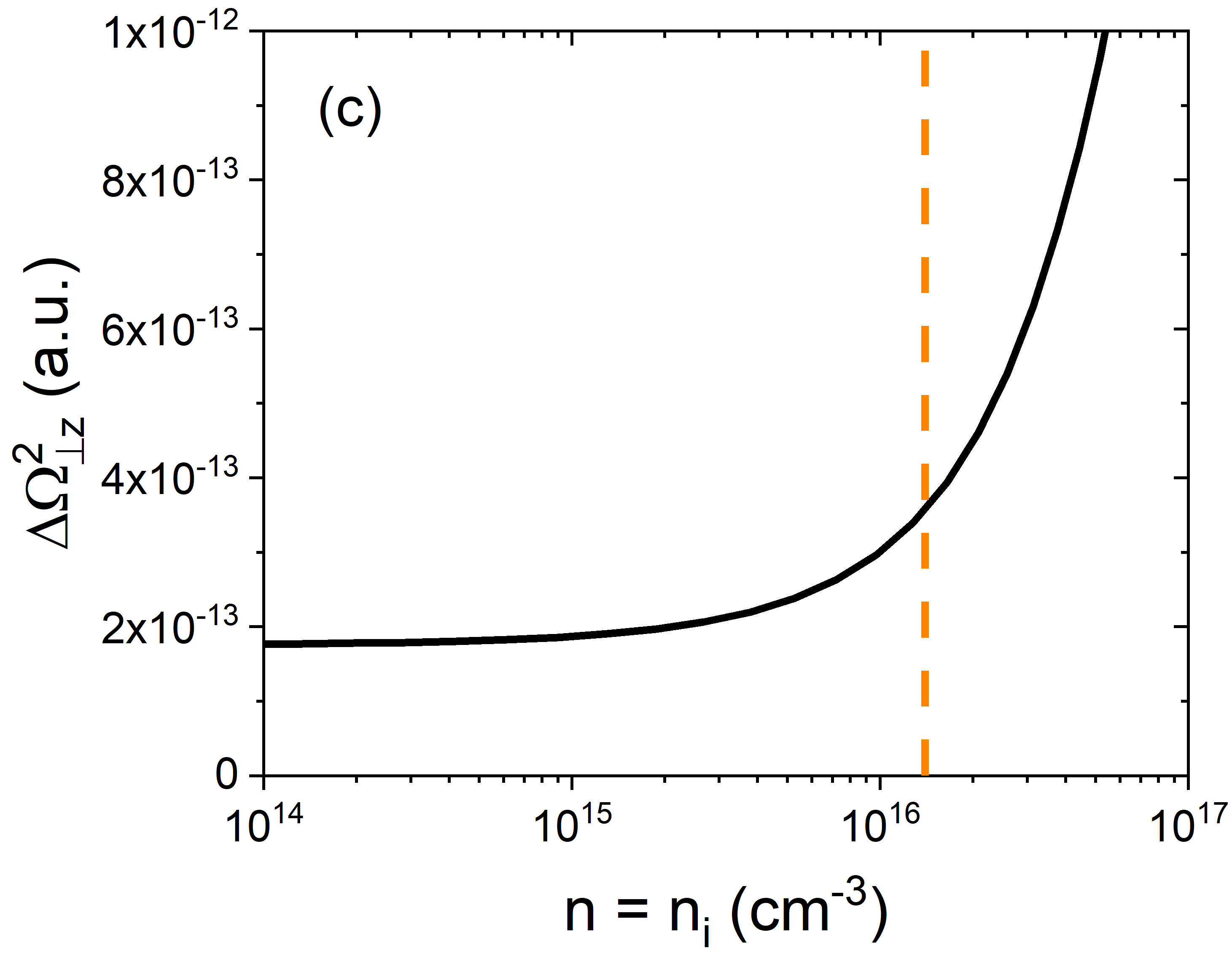} \caption{(a) $\tau_{s}$ and (b) $\tau_{p}$ of $n$-GaAs with different doping
concentrations $n_{i}$ at 30 K with different scattering mechanisms.
``All'' represents all the e-ph, e-i and e-e scattering mechanisms
being considered. (c) $\Delta\Omega_{\perp z}^{2}$ as a function
of carrier density $n$, where ${\bf \Delta\Omega_{\perp z}}$ is
the fluctuation amplitude of Larmor frequency defined in Eq. \ref{eq:fluctuation_Larmor}~\citep{xu2021ab}.\label{fig:doping}}
\end{figure*}

Figure~\ref{fig:doping} shows $\tau_{s}$ and $\tau_{p}$ with different
$n_{i}$ at 30 K with individual and total scattering pathways, respectively.
It is found that the roles of different scattering mechanism differ
considerably between spin and carrier relaxation processes. Specifically,
for the carrier relaxation in Fig.~\ref{fig:doping}b, except when
$n_{i}$ is very low (e.g. at $10^{14}$ cm$^{-3}$), the e-i scattering
dominates. On the other hand, for the spin relaxation in Fig.~\ref{fig:doping}a,
the e-e scattering dominates except at very high concentration (above
$10^{17}$ cm$^{-3}$), while e-i scattering is only important in
the very high doping region (close to or above $10^{17}$ cm$^{-3}$).

Figure~\ref{fig:doping} shows the calculated $\tau_{s}$ has a maximum
at $n_{i}\,=\,1\text{-}2\times10^{16}$ cm$^{-3}$, and $\tau_{s}$
decreases fast with $n_{i}$ going away from its peak position. This
is in good agreement with the experimental finding in Ref. \citenum{kikkawa1998resonant},
which also reported $\tau_{s}$ at $n_{i}=10^{16}$ cm$^{-3}$ is
longer than $\tau_{s}$ at other lower and higher $n_{i}$ at a low
temperature (a few Kelvin). The $n_{i}$ dependence of $\tau_{s}$
may be qualitatively interpreted from the DP relation\citep{vzutic2004spintronics}
$\tau_{s,z}\sim\tau_{p}\Delta\Omega_{\perp z}^{2}$ (Eq. \ref{eq:DP}),
where ${\bf \Delta\Omega_{\perp z}}$ is the fluctuation amplitude
of Larmor frequency defined in Eq. \ref{eq:fluctuation_Larmor}. From
Fig.~\ref{fig:doping}, we find that with $n_{i}$ from $10^{14}$
cm$^{-3}$ to $5\times10^{15}$ cm$^{-3}$, $\tau_{p}$ decreases
rapidly (black curve in Fig.~\ref{fig:doping}b) and $\Delta\Omega_{\perp z}^{2}$
remains flat in Fig.~\ref{fig:doping}c, which may explain why $\tau_{s}$
increases in Fig.~\ref{fig:doping}a based on the DP relation; however,
when $n_{i}>10^{16}$ cm$^{-3}$, $\tau_{p}$ decreases with a similar
speed but $\Delta\Omega_{\perp z}^{2}$ experiences a sharp increase,
which may explain why $\tau_{s}$ decreases in Fig.~\ref{fig:doping}b
and owns a maximum at $10^{16}$ cm$^{-3}$.

Note that although the DP relation is intuitive to understand the
cause of doping-level dependence of spin lifetime, it may break down
when we evaluate individual scattering processes. For example, when
$n_{i}$ increases from $10^{14}$ cm$^{-3}$ to $10^{15}$ cm$^{-3}$,
both carrier lifetime $\tau_{s}$ and $\tau_{s,z}$ due to e-i scattering
decrease while the internal magnetic field remains unchanged. Moreover,
the simple empirical relation cannot possibly explain our first-principles
results that the e-e and e-i scatterings have largely different contributions
in carrier and spin relaxation. First-principles calculations are
critical to provide unbiased mechanistic insights to spin and carrier
relaxation of general systems.

\subsubsection{Graphene on hBN}

Graphene samples exhibit exciting spintronic properties such as long
$\tau_{s}$ and $l_{s}$ at room temperature \citep{avsar2020colloquium,drogeler2016spin}.
In practice, graphene is usually supported by a substrate\citep{dean12010GraphhBNMobIncrease,han2011spin,guimaraes2014controlling,drogeler2016spin,raes2016GraphSiO2AnisObliqePrecession}.
Actually $\tau_{s}$ of the free-standing graphene sample was found
relatively low\citep{guimaraes2012spin} $\sim$150 ps at 300 K (compared
with the longest reported value $\sim$12 ns), probably due to free-standing
graphene samples often have more imperfections. Therefore, it is important
to understand spin relaxation in supported graphene.

\begin{figure}[ht!]
\includegraphics[width=1\columnwidth]{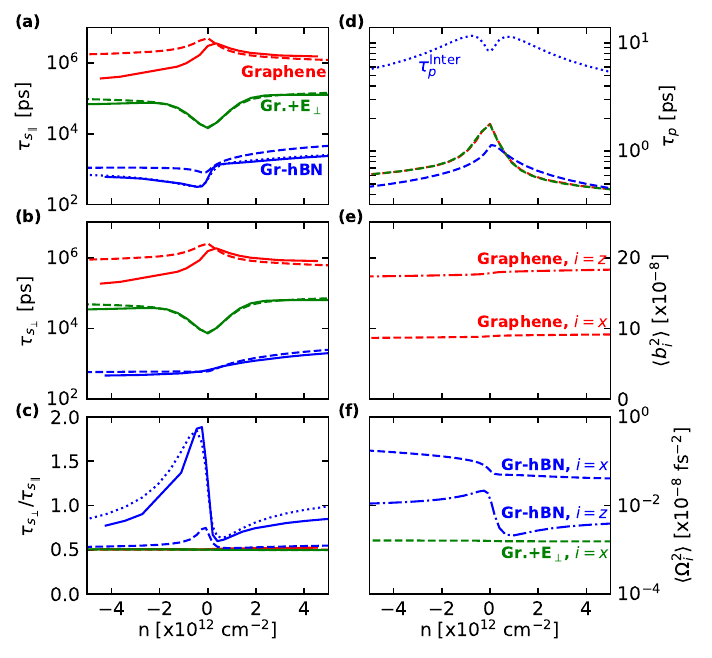} \caption{Theoretical spin and carrier relaxation in graphene (red lines), graphene
at $E_{\perp}$=0.4 V/nm (Gr+$E_{\perp}$, green lines) and graphene
on hBN (Gr-hBN, blue lines) as a function of $n$ (positive or negative
for electron and hole doping respectively) at room temperature. (a)
In-plane spin lifetime $\tau_{s\parallel}$, (b) out-of-plane spin
lifetime $\tau_{s\perp}$ and (c) their ratios $\tau_{s\perp}/\tau_{s\parallel}$
calculated using FPDM approach (solid lines) compared with those estimated
from EY/DP models with first-principles inputs (dashed for conventional
DP and dotted for modified DP with intervalley scattering contribution
lines). (d) Total $\tau_{p}$ (three dashed lines) and intervalley
only contribution for Gr-hBN (dotted blue line), (e) $\left\langle b_{i}^{2}\right\rangle $
and (f) $\left\langle \Omega_{i}^{2}\right\rangle $, all predicted
from first-principles for the DP and EY models estimates of $\tau_{s}$
in (a-c)~\citep{habib2022electric}. \label{fig:tauCarrComp}}
\end{figure}

From Fig.~\ref{fig:tauCarrComp}, we find that the EY model for graphene,
and the DP model for Gr+$E_{\perp}$ and Gr-hBN, agree qualitatively
with FPDM predictions, but with some important quantitative differences
discussed next. First, the EY model for graphene is more accurate
for electrons than for holes, for both $\tau_{s\parallel}$ and $\tau_{s\perp}$
(Fig.~\ref{fig:tauCarrComp}(a-b)). The conventional DP model matches
FPDM predictions quantitatively for both $\tau_{s\parallel}$ and
$\tau_{s\perp}$ of Gr+$E_{\perp}$, but only for $\tau_{s\perp}$
of Gr-hBN.

The discrepancy of the conventional DP model for $\tau_{s\parallel}$
of Gr-hBN can be rectified by modifying the model. Briefly, the DP
model assumes that ${\bf B}^{\mathrm{in}}$ effectively changes randomly
each time the electron scatters. The in-plane magnetic field ${\bf B}_{||}^{\mathrm{in}}$
rotates over the Fermi circle and covers all in-plane directions,
satisfying this condition, in both Gr+$E_{\perp}$ and Gr-hBN. However,
the out-of-plane magnetic field ${\bf B}_{\perp}^{\mathrm{in}}$,
which matters only for $\tau_{s\parallel}$ and is present only for
Gr-hBN, has the same direction within each valley. Consequently, only
intervalley scattering will change the ${\bf B}_{\perp}^{\mathrm{in}}$
for a given electron spin. As proposed in Ref.~\citenum{cummings2017giantspinlifetime},
this can be captured by changing the DP model from $(\tau_{s,x}^{\mathrm{DP}})^{-1}\sim\tau_{p}\left(\Delta{\bf \Omega}_{\perp x}\right)^{2}=\tau_{p}\left\langle \Omega_{y}^{2}+\Omega_{z}^{2}\right\rangle $
as given by Eq.~\ref{eq:DP} for in-plane $x$ spins, to 
\begin{equation}
\left(\tau_{s,x}^{\mathrm{mDP}}\right)^{-1}\approx\tau_{p}\left\langle \Omega_{y}^{2}\right\rangle +\tau_{p}^{\mathrm{Inter}}\left\langle \Omega_{z}^{2}\right\rangle ,\label{eq:ModifiedDP}
\end{equation}
where $\tau_{p}^{\mathrm{Inter}}$ is the intervalley scattering time
(dotted line in Fig.~\ref{fig:tauCarrComp}(d)). This modified DP
model agrees with FPDM predictions for $\tau_{s\parallel}$ of Gr-hBN
(Fig.~\ref{fig:tauCarrComp}(a)).

The ratio $\tau_{s\perp}/\tau_{s\parallel}$ (Fig.~\ref{fig:tauCarrComp}(c))
is nearly 1/2 for graphene, consistent with the EY relation Eq. \ref{eq:taus_b2_relation}
and the fact that $\left\langle b_{||}^{2}\right\rangle /\left\langle b_{\perp}^{2}\right\rangle $
is also 1/2 (Fig.~\ref{fig:tauCarrComp}(e)). This ratio remains
unchanged for Gr+$E_{\perp}$, but now because $\left(\Delta{\bf \Omega}_{\perp z}\right)^{2}=\langle\Omega_{x}^{2}+\Omega_{y}^{2}\rangle=2\langle\Omega_{y}^{2}\rangle$,
while $\left(\Delta{\bf \Omega}_{\perp x}\right)^{2}=\langle\Omega_{y}^{2}\rangle$
since $\Omega_{z}=0$ (Fig.~\ref{fig:tauCarrComp}(f)), leading to
$\tau_{s,x}^{\mathrm{DP}}=2\tau_{s,z}^{\mathrm{DP}}$ using Eq.~\ref{eq:DP}.
This ratio deviates substantially from 1/2 only for Gr-hBN (Fig.~\ref{fig:tauCarrComp}(c))
due to the substrate-induced $\Omega_{z}\ne0$. The conventional DP
model (Eq.~\ref{eq:DP}) only captures part of this dramatic effect
seen in the FPDM calculations, while the modifications in Eq.~\ref{eq:ModifiedDP}
account for $\Omega_{z}$ correctly and agree with the FPDM results
in Fig.~\ref{fig:tauCarrComp}(c).

$\tau_{s}$ decrease with increasing carrier density magnitude in
graphene (Fig.~\ref{fig:tauCarrComp}(a,b)), but this trend reverses
for both inversion-symmetry-broken cases, in agreement with some experiments.\citep{drogeler2016spin,zomer2012ChBNlongdistSpinhighmob}
The overall $\tau_{s}$ are reduced by one-two orders of magnitude
in free-standing graphene by $E_{\perp}$ of 0.4~V/nm, down from
$\mu s$ to tens of ns. $\left\langle \Omega_{i}^{2}\right\rangle $
of Gr-hBN in Fig.~\ref{fig:tauCarrComp}(f) is about 100 times larger
than that of Gr+$E_{\perp}$, further reducing $\tau_{s}$ of Gr-hBN
to the ns scale, comparable to experimental measurements.\citep{guimaraes2014controlling,kamalakar2015long,han2011spin,drogeler2016spin}

Finally, $\tau_{s}$ is mostly symmetric between electrons and holes
for Gr+$E_{\perp}$ (Fig.~\ref{fig:tauCarrComp}(a, b)). However,
we find hole $\tau_{s}$ to be typically 2-3 times smaller than electrons
for both free-standing graphene and Gr-hBN. On hBN, this asymmetry
is captured by the (modified) DP model and stems primarily from the
larger spin-splitting and hence $\left\langle \Omega_{i}^{2}\right\rangle $
in the valence band compared to the conduction band (not shown), consistent
with previous calculations.\citep{zollner2019GraphhBNAbInitioSpinRelax}
Importantly, this effect depends sensitively on the substrate, and
even on hBN, could reverse for a different layer stacking.\citep{zollner2019GraphhBNAbInitioSpinRelax}
Consequently, experiments may find electron-hole asymmetries of either
sign depending on the substrate and precise structure,\citep{han2011spin,zomer2012ChBNlongdistSpinhighmob,guimaraes2014controlling,kamalakar2015long,drogeler2016spin,raes2016GraphSiO2AnisObliqePrecession,zollner2019GraphhBNAbInitioSpinRelax}
and we focus here on the comparison between FPDM predictions and DP
model for the specific lowest-energy stacking of Gr-hBN.

\subsection{Spin dephasing under magnetic field}

As introduced in Sec. \ref{subsec:relaxation-and-dephasing}, the
dephasing time (or transverse time) of a spin ensemble is called $T_{2}^{*}$
and describes the decay of the total excess spin ${\bf S}-{\bf S}^{\mathrm{eq}}$
at an nonzero ${\bf B}^{\mathrm{ext}}$ perpendicular to ${\bf S}-{\bf S}^{\mathrm{eq}}$.
As discussed in Sec. \ref{par:theory-Bfield}, $T_{2}^{*}$ can be
simulated straightforwardly through FPDM simulations with Zeeman Hamiltonian
considering both spin and orbital angular momenta (Eq. \ref{eq:Zeeman}
and \ref{eq:Lmatrix}). For the purpose of analysing and understanding
spin dephasing, one key parameter is the Land\'e $g$-factor $\widetilde{g}$
(Eq. \ref{eq:gfac}). It value relates to ${\bf B}^{\mathrm{ext}}$-induced
energy splitting (Zeeman effect) $\Delta E\left({\bf B}^{\mathrm{ext}}\right)$
and Larmor precession frequency $\Omega$, satisfying $\Omega\approx\Delta E=\mu_{B}B^{\mathrm{ext}}\widetilde{g}$.
More importantly, the $g$-factor fluctuation (near Fermi surface
or $\mu_{F,c}$) $\Delta\widetilde{g}$ (Eq. \ref{eq:g_sigma}) leads
to a nonzero $\Delta\Omega$ (Eq. \ref{eq:dg_induced_dOmega}) and
then $T_{2}^{*}$ due to DP or FID mechanism.

Spintronics in halide perovskites has drawn significant attention
in recent years, due to highly tunable spin-orbit fields and intriguing
interplay with lattice symmetry. Recently, we simulate\citep{xu2023spin}
$\widetilde{g}$, $\Delta\widetilde{g}$ and $T_{2}^{*}$ of a typical
halide perovskite - CsPbBr$_{3}$. We find that $T_{2}^{*}$ is sensitive
to $B^{\mathrm{ext}}$ at $T<$20 K but not at $T\ge$20 K. At 4 K,
we predict that $\left(T_{2}^{*}\right)^{-1}$ is linear to $B^{\mathrm{ext}}$
at $B^{\mathrm{ext}}\ge$0.4 Tesla and the predicted slope of $\left(T_{2}^{*}\right)^{-1}$
is in good agreement with experimental data. Moreover, we predict
strong $n$-dependence of $\widetilde{g}$, $\Delta\widetilde{g}$
and $T_{2}^{*}$. Together with FPDM simulations of spin relaxation
time $T_{1}$ of CsPbBr$_{3}$ at various conditions, our work provides
fundamental insights on how to control and manipulate spin relaxation/dephasing
in halide perovskites, which are vital for their spintronics and quantum
information applications.

\section{Outlooks}



The first-principles density-matrix dynamics (FPDM) approach is an important technique
for studying spin and electron dynamics and transport, with outstanding
advantages: i) it can accurately describe various interactions, scattering
processes, and spin precession simultaneously; ii) it can describe processes far from
equilibrium; iii) it can simulate dynamical processes on various time scales
from femtoseconds to milliseconds. We have shown its success in simulating
ultrafast spin dynamics, spin and charge transport, relaxation and dephasing. In the future study, the FPDM approach still has a broad open area for new theory development.

\subsubsection{Exciton dynamics}

Exciton is a bound electron-hole pair generated by optical excitation. Excitons play an important role in optical properties of semiconductors, in particular in low-dimensional systems. Exciton dynamics has been extensively studied using different methods, in particular non-equilibrium Greens' function theory (NEGF) within Kadanoff-Baym equations\cite{paleari2022exciton,chan2021giant,molina2017ab}.
We focus on theoretical simulations of exciton spin relaxation based on Lindbladian DM master equation with quantum description of the exciton-phonon scattering.
Understanding the exciton spin relaxation is also important
for understanding optical measurements of spin dynamics. By replacing
electrons to excitons, the existing FPDM approach can be generalized
to simulation spin dynamics of excitons. The exciton dynamics can
involve many types of processes including exciton-light interaction,
exciton-phonon scattering, phonon-mediated exciton recombination and dissociation,
exciton--exciton annihilation, etc. Initially, we will focus on exciton-light
interaction and exciton-phonon scattering, which are two most important processes leading to exciton spin relaxation.

The exciton-light interaction is responsible to exciton spin generation
by light absorption with a circularly-polarized pump pulse and exciton
spin relaxation through exciton-phonon scattering and exciton radiative recombination. The light
absorption for excitons is similar to that for electrons (see Sec.
\ref{par:laser}) except that the momentum matrix elements are now
between two excitonic states~\citep{Guo_2021, C9TC02214G,Smart2021-ef}. 


The exciton-phonon scattering is responsible to exciton spin relaxation.
According to Refs.~\citenum{Antonius2022, chen2020exciton}, the exciton-phonon scattering
matrix elements can be approximately obtained from the excition wavefunction at finite momentum and electron-phonon
scattering matrix elements. Recently we implemented the exciton-phonon scattering matrix elements and used them to
simulate phonon-assisted indirect exciton radiative recombination.
In the density-matrix master equation, the term of the exciton-phonon
scattering is rather similar to the electron-phonon scattering, except
that the excitons are bosons while the electrons are fermions. Therefore,
the implementation of the exciton-phonon scattering in the frame of
FPDM approach can be straightforward.

\subsubsection{Circular photogalvanic effect (CPGE)}
CPGE is the effect that under the circularly-polarized light a DC
current may be induced in a solid-state material, and widely presents
in materials without inversion symmetry, in absence of p-n junction and applied electric field. In recent years, CPGE has
attracted growing interests in the fields of topological physics and spin-optotronics~\citep{Watanabe2021,De_Juan2017-ze}.   
In order to deeply understand CPGE,  theoretical methods have
been developed based on perturbation theory. However, several critical issues remain: (i) the scattering mechanisms are highly
simplified with a single relaxation time approximation; (ii) spontaneous recombination was rarely considered; (iii) available theory requires different formulation for each contribution i.e. interband or intraband contributions; (iv) many-body effects  were not considered. These issues may be resolved based on our FPDM approach.

If a constant laser field is applied to an inversion symmetry broken system, the quantum master equation for describing density matrix of the electronic system reads:

\begin{align}
\frac{d\rho}{dt}= & \frac{d\rho}{dt}|_{\mathrm{laser}}+\frac{d\rho}{dt}|_{\mathrm{sp-em}}+\frac{d\rho}{dt}|_{\mathrm{e-ph}},
\end{align}

where $\frac{d\rho}{dt}|_{\mathrm{laser}}$ is coherent dynamics due
to a laser field (Eq. \ref{eq:laser-dynamics}), $\frac{d\rho}{dt}|_{\mathrm{sp-em}}$
and $\frac{d\rho}{dt}|_{\mathrm{e-ph}}$ are the spontaneous emission
(Eq. \ref{eq:sp-em}) and e-ph scattering terms (Eq. \ref{eq:scattering_BornMarkov}
and \ref{eq:Peph}), respectively.

CPGE is measuring the steady state current when a nonmagnetic system is under a constant circularly-polarized laser field for a long enough time. In the velocity gauge,
the current density is computed using\citep{ventura2017gauge}

\begin{align}
{\bf J}= & \mathrm{Tr}\left(\rho{\bf j}\right),\\
{\bf j}_{kmn}\left(t\right)= & -e\left({\bf v}_{kmn}+\delta_{mn}\frac{e}{m_{e}}{\bf A}\left(t\right)\right),
\end{align}

where ${\bf v}$ is the velocity operator matrix and ${\bf v}={\bf p}/m_{e}$.
The average value of the second term of ${\bf j}$ is usually zero
so that this term does not contribute to CPGE.

\subsubsection{More on scattering terms}

Although in our FPDM approach, the \emph{ab-initio} treatment of
quantum scattering is rather general, the following theoretical development
can further improve the applicability of our methods to different systems:
\\
\paragraph{Anharmonic phonons.}

In current implementation, phonons are harmonic and it is required
that the material must be dynamically stable at zero temperature.
However, for some soft materials (e.g., hybrid halide perovskites), significant
anharmonic effects appear at high-$T$ phase, which requires extracting phonon properties from finite-temperature simulations, e.g. using \emph{ab-initio} molecular dynamics. Such approach has been implemented~\cite{hellman2013temperature} for studying phonon, electron-phonon, and carrier transport properties, which can be also applied to study the anharmonic effect on spin relaxation under our FPDM framework.   
\\

\paragraph{Fr\"ohlich interaction and LO-TO splitting in doped semiconductor.}

The intraband dielectric screening is not considered for Fr\"ohlich
interaction and LO-TO splitting in current implementation. This may
be problematic in doped semiconductor and may lead to significant
errors at moderate or high carrier density. Since we have already
implemented the intraband dielectric function (Eq. \ref{eq:dielectric-intra}),
its effect can be straightforwardly included in FPDM approach similar
to Ref. \citenum{macheda2022electron}.
\\
\paragraph{Short-range contribution to the electron-ionized-impurity scattering.}

For the electron-ionized-impurity scattering, we currently only consider
the long-range contribution. This is appropriate for certain systems
like GaAs, but may be problematic when intravalley processes dominate
spin relaxation or the short-range spin-flip electron-ionized-impurity
scattering is unimportant to spin relaxation. Therefore, it is important
to include both long-range and short-range contributions to the electron-ionized-impurity
scattering. Similar to Ref. \citenum{verdi2015frohlich}, the long-range
and short-range parts are treated separately - the short-range part
is treated from first principles similar to neutral impurity, and the
long-range part is simulated using the screened Coulomb potential.
\\

\section*{Acknowledgements}

This work is supported by National Science Foundation
under grant No. DMR-1956015. 
This research used resources of the Center for Functional
Nanomaterials, which is a US DOE Office of Science Facility, and the
Scientific Data and Computing center, a component of the Computational
Science Initiative, at Brookhaven National Laboratory under Contract
No. DE-SC0012704, the lux supercomputer at UC Santa Cruz, funded by
NSF MRI grant AST 1828315, the National Energy Research Scientific
Computing Center (NERSC) a U.S. Department of Energy Office of Science
User Facility operated under Contract No. DE-AC02-05CH11231, the Extreme
Science and Engineering Discovery Environment (XSEDE) which is supported
by National Science Foundation Grant No. ACI-1548562 \citep{xsede}.

\bibliographystyle{apsrev4-1}
\bibliography{ref}

\end{document}